%% file: ms.tex
\documentclass[iop]{emulateapj}
\usepackage{lscape}

\input{newcom}
\newcommand{\hsten}{HS10}
\newcommand{\hs}{HS12}
\newcommand{\hskjm}{HSKJM}

\newcommand{\Fsource}{\ensuremath{F_\mathrm{X}^{0.5-2.0}}}
\newcommand{\pvalue}{\ensuremath{p~\mathrm{value}}}
\newcommand{\Stotal}{\ensuremath{S_{0.5-2.0}}}
\newcommand{\twoMK}{\ensuremath{(\mbox{2--3}) \times 10^6~\K}}
\newcommand{\tenMK}{\ensuremath{10 \times 10^6~\K}}

\shorttitle{GALACTIC HALO X-RAY EMISSION}
\shortauthors{HENLEY \& SHELTON}

\begin{document}

\title{An \textit{XMM-Newton} Survey of the Soft X-ray Background. III.
       The Galactic Halo X-ray Emission}
\author{David B. Henley and Robin L. Shelton}
\affil{Department of Physics and Astronomy, University of Georgia, Athens, GA 30602; dbh@physast.uga.edu}

\begin{abstract}
We present measurements of the Galactic halo's X-ray emission for 110 \xmm\ sight lines, selected to
minimize contamination from solar wind charge exchange emission.
We detect emission from few million degree gas on $\sim$4/5 of our sight lines. The temperature is
fairly uniform ($\mbox{median} = 2.22 \times 10^6~\K$, $\mbox{interquartile range} = 0.63 \times
10^6~\K$), while the emission measure and intrinsic 0.5--2.0~\kev\ surface brightness vary by over
an order of magnitude ($\sim$$(\mbox{0.4--7}) \times10^{-3}~\emismeas$ and $\sim$$(\mbox{0.5--7})
\times10^{-12}~\flux\ \pdegsq$, respectively, with median detections of $1.9 \times
10^{-3}~\emismeas$ and $1.5 \times 10^{-12}~\flux\ \pdegsq$, respectively).
The high-latitude sky contains a patchy distribution of few million degree gas. This gas exhibits a
general increase in emission measure toward the inner Galaxy in the southern Galactic hemisphere.
However, there is no tendency for our observed emission measures to decrease with
increasing Galactic latitude, contrary to what is expected for a disk-like halo morphology.
The measured temperatures, brightnesses, and spatial distributions of the gas can be used to place
constraints on models for the dominant heating sources of the halo.  We provide some discussion of
such heating sources, but defer comparisons between the observations and detailed models to a later
paper.
\end{abstract}

\keywords{Galaxy: halo ---
  ISM: structure ---
  X-rays: diffuse background ---
  X-rays: ISM}

\section{INTRODUCTION}
\label{sec:Introduction}

Observations of the $\sim$0.1--1~\kev\ diffuse soft X-ray background (SXRB; e.g.,
\citealt{mccammon90}) show that $\sim$$(\mbox{1--3}) \times 10^6~\K$ plasma is present in the halo
of the Milky Way
\citep{burrows91,wang95,pietz98,wang98,snowden98,snowden00,kuntz00,smith07a,galeazzi07,henley08a,lei09,yoshino09,gupta09b,henley10b}.
The presence of this hot plasma is confirmed by the observation of zero-redshift \OVII\ and
\OVIII\ absorption lines in the X-ray spectra of active galactic nuclei (AGN;
\citealt{nicastro02,fang03,rasmussen03,mckernan04,williams05,fang06,bregman07,yao07a,yao08,hagihara10,sakai12,gupta12}). The
extent and mass of this hot gas is disputed: \citet{gupta12} argue that its extent is
$\ga$100~\kpc\ and that it contains a significant fraction of the Galaxy's baryonic mass (see also
\citealt{fang13}), while \citet[and references therein]{wang12} argue that the scale height of the
hot gas is only a few kpc, in which case it would contribute a negligible amount to the Galaxy's
baryons \citep{fang13}.

The origin of this hot halo gas is uncertain. Two main processes are thought to play a role in
heating the halo. The first is supernova (SN) driven outflows from the Galactic disk
\citep[e.g.,][]{shapiro76,bregman80,norman89,joung06}. In such an outflow, the material may
subsequently fall back to the disk, in a so-called galactic fountain. The second process is
accretion of material from the intergalactic medium
\citep[e.g.,][]{toft02,rasmussen09,crain10}. However, the relative importance of these two processes
is not well known.

\citet[hereafter \hskjm]{henley10b} tested models of the hot halo gas using a sample of 26 SXRB
spectra extracted from archival \xmm\ observations between $l = 120\degr$ and
240\degr\ \citep[hereafter \hsten]{henley10a}. They compared the observed X-ray temperatures and
emission measures of the hot halo with the distributions expected from different physical models.
\hskjm's analysis favored fountains of hot gas \citep{joung06} as a major, possibly dominant,
contributor to the halo X-ray emission in the \xmm\ band over extraplanar supernova remnants
\citep{shelton06}. However, in the absence of X-ray surface brightness predictions from disk galaxy
formation models, they were unable to rule out the possiblity that an extended halo of accreted
material also contributed to the observed emission \citep{crain10}.

Here, we expand upon \hskjm's observational analysis, analyzing $\sim$4 times as many sight lines.
Our observations are drawn from a new \xmm\ SXRB survey which spans the full range of Galactic
longitudes \citep[hereafter \hs]{henley12b}, and which supersedes the \hsten\ survey from which the
\hskjm\ sample was drawn. As in \hskjm, our observations were selected because they should be less
affected by solar wind charge exchange (SWCX) emission \citep{cravens00,robertson03a,koutroumpa06}
-- time-variable X-ray line emission which arises within the solar system from charge exchange
reactions between solar wind ions and neutral H and He
(\citealt{cravens01,wargelin04,snowden04,koutroumpa07,fujimoto07,kuntz08a,henley08a}; \hsten;
\citealt{carter08,carter10,carter11,ezoe10,ezoe11}).  In a separate paper, we will use these
observations to test models of the hot halo gas (D.~B. Henley et~al., 2013, in preparation).

The remainder of this paper is organized as follows. In Section~\ref{sec:Observations} we describe
the observation selection and data reduction. In Section~\ref{sec:Method} we describe our spectral
analysis method. We present the results in Section~\ref{sec:Results}. We discuss and summarize our
results in Sections~\ref{sec:Discussion} and \ref{sec:Summary}, respectively.

\section{OBSERVATIONS}
\label{sec:Observations}

\subsection{Observation Selection}
\label{subsec:ObservationSelection}

The observations that we analyze here are a subset of those analyzed by \hs, who extracted SXRB
\OVII\ and \OVIII\ intensities from 1880 archival \xmm\ observations spread across the sky.  In
order to minimize SWCX contamination, we apply various filters to the data (\hskjm). In particular,
to minimize contamination from geocoronal SWCX and near-Earth heliospheric SWCX, we only use the
portions of the \xmm\ observations during which the solar wind proton flux was low or moderate.  If
excising the periods of high solar wind proton flux from an \xmm\ observation resulted in too
little usable observation time, the observation was rejected (see Section~2.4 of \hs). After this
solar wind proton flux filtering, 1003 observations are usable (\hs, Table~2).  We apply additional
filters to these observations as follows.  We minimize heliospheric SWCX contamination by using only
observations toward high ecliptic latitudes ($|\beta| > 20\degr$) taken during solar minimum (after
00:00UT on 2005 Jun 01\footnote{This date, taken from \hs, was estimated using sunspot data from the
  National Geophysical Data Center (http://www.ngdc.noaa.gov/stp/SOLAR/). Note that this date is
  later than the one used in \hskjm, as \hs\ defined an ``Intermediate'' phase of the solar cycle
  between solar maximum and solar minimum. We did not define an end date for the solar minimum
  phase, as the sunspot data imply that this phase lasted at least until the most recent observation
  in the \hs\ catalog (carried out on 2009 Nov 03--04).}). As we are interested in the Galactic
halo, we use only observations toward high Galactic latitudes ($|b| > 30\degr$), and exclude
observations toward the Magellanic Clouds, the Eridanus Enhancement \citep{burrows93,snowden95b},
and the Scorpius-Centaurus (Sco-Cen) superbubble \citep{egger95}. Note that although we do not
explicitly exclude the observations identified as being SWCX-contaminated by
\citet[Table~A.1]{carter11}, none of these observations are in our final sample.

The above criteria result in 163 observations being selected from \hs's original set of 1003.  The
observation IDs, names of the original targets,\footnote{In general, the target names were obtained
  from the FITS file headers. If the target name was abbreviated or truncated, we attempted to
  determine the full name of the target from SIMBAD (http://simbad.u-strasbg.fr/simbad/). For a
  small number of targets, we were unable to determine the full name.} and pointing directions for
these 163 observations are shown in Columns~2 through 5 of Table~\ref{tab:FitResults} (Columns~6
through 9 contain additional observation information [Section~\ref{subsec:DataReduction}] and
Columns~10 through 15 contain the spectral fit results [Section~\ref{sec:Results}]). If the original
target was a bright X-ray source, we excised it from the data, since our goal is to measure the
diffuse SXRB emission in each \xmm\ field (see Section~\ref{subsec:DataReduction}, below). Note that
these 163 observations represent fewer than 163 different sight lines. If a set of observations are
separated by less than 0.1\degr, we group them into a single sight line, and then fit our spectral
model (Section~\ref{subsec:ModelDescription}) to all the observations simultaneously. In such cases,
the observations for a given sight line are listed in the table on and below the row containing the
sight line number (e.g., the results for sight line 20 were obtained by simultaneously fitting to
the spectra from observations 0400920201 and 0400920101).

Our set of 163 observations includes a cluster of 28 observations near $(l,b) \approx
(326\degr,-58\degr)$. These observations represent 27 different sight lines, which we have numbered
103.1 through 103.27 (sight line 103.8 consists of two observations). In order to avoid oversampling
this region of the sky in our subsequent analysis, we treat these 27 sight lines as a single
sight line, whose halo temperature and emission measure are found from the weighted means of the halo
temperatures and emission measures of the individual sight lines. We tabulate these mean values as
the results for sight line 103 in columns~12 and 13 of Table~\ref{tab:FitResults}. Similarly, the
Galactic coordinates for this sight line are the means of the longitudes and latitudes for the
individual sight lines. The subsequent analysis will use the mean results for sight line 103.

After grouping together observations of the same sight line, and combining the results from
sight lines 103.1 through 103.27 as described above, our set of 163 observations yields 110
measurements of the halo's temperature and emission measure. The locations of our sight lines on the
sky are shown in Figure~\ref{fig:Maps}, below.

Note that our sample of observations includes 20 of the 26 observations analyzed in \hskjm. Of the
remaining six observations, five (0200960101, 0303260201, 0303720201, 0303720601, 0306370601) are
excluded due to our using a later date to define the beginning of the solar minimum phase. The sixth
observation (0305290201) is not included in \hs's catalog, and so is not included here, as it
exhibits strong residual soft proton contamination (see Section~3.5 of \hs). Observations 0306060201
and 0306060301 were analyzed independently in \hskjm, but here they are grouped together (sight line
43).

\subsection{Data Reduction}
\label{subsec:DataReduction}

The data reduction is described in Section~2 of \hs\ (see also Section~3 of
\hsten). Here, we give an overview of the process. The data reduction was carried out
using the \xmm\ Extended Source Analysis
Software\footnote{http://heasarc.gsfc.nasa.gov/docs/xmm/xmmhp\_xmmesas.html} (\esas;
\citealt{kuntz08a,snowden11}), as included in version 11.0.1 of the \xmm\ Science Analysis
Software\footnote{http://xmm.esac.esa.int/sas/} (SAS).  Note that we re-extracted all the EPIC-MOS
spectra from scratch for the current analysis, using a lower source removal flux threshold than in
\hs\ (see below).

Each observation was first processed with the SAS \texttt{emchain} script to produce a calibrated
events list for each exposure. Then, the \esas\ \texttt{mos-filter} script was used to identify and
excise periods within each exposure that were affected by soft proton flaring. As indicated above,
periods of high solar wind proton flux ($> 2 \times 10^8$~\pcmsq\ \ps) were also removed from the
data. The usable MOS1 and MOS2 exposure times that remain after this filtering are shown in
Columns~6 and 8 of Table~\ref{tab:FitResults}, respectively.

Because our goal is to measure the diffuse Galactic halo emission, we removed bright sources from
the \xmm\ data. As described in \hsten\ (Section~3.3) and \hs\ (Section~2.2), we identified
and removed bright and/or extended sources that would not be adequately removed by the automated
source removal (described below). If the source to be removed was the original observation target,
we centered the exclusion region on the target's coordinates; otherwise, the exclusion region was
positioned by eye. In all cases we used circular exclusion regions. We chose the radii of these
regions by eye, although in some cases we used surface brightness profiles to aid us. As noted in
\hsten\ and \hs, we erred on the side of choosing larger exclusion regions, at the expense
of reducing the number of counts in the SXRB spectra.

In general, we used the same source exclusion regions that we used in \hsten\ and \hs.
These were chosen from a visual inspection of broadband X-ray images, which had undergone only the
basic processing described above. One change we made was in our processing of observation 0305860301
(sight line 100). In \hs, we did not exclude the target galaxy, NGC~300, as a visual inspection of
the X-ray images suggested it would not significantly contaminate the SXRB measurements. Here,
however, we decided to err on the side of caution and excised the galaxy from the X-ray data before
extracting the SXRB spectrum.

After our initial spectral extraction (described below), we found that on nine of our sight lines
our spectral fitting (Section~\ref{sec:Method}) yielded halo temperatures of $\sim$\tenMK.  Such
temperatures are much higher than those that are typically observed in the halo ($T \sim \twoMK$;
\citealt{smith07a,galeazzi07,gupta09b,lei09,yoshino09}; \hskjm). For these sight lines, we
re-examined the X-ray images. In particular, we used \esas\ tools to create adaptively smoothed,
particle-background-subtracted, flat-fielded images in the 0.4--1.3~\kev\ band (the upper energy
limit was chosen to avoid the Al instrumental line at 1.49~\kev). These images revealed regions of
diffuse emission that had not been adequately excised by the original exclusion regions. We
therefore excluded these additional regions and re-extracted the spectra. The affected observations
are indicated by a ``d'' in Column~1 of Table~\ref{tab:FitResults}. After this change, our spectral
analysis yields halo temperatures of $\sim$\twoMK\ on all but one of these nine sight lines.  Note that
the additional sources we have removed appear not to have contaminated the $\sim$\twoMK\ halo
emission. Therefore, the measurements of this emission from the other sight lines (for which we used
the source exclusion regions straight from \hsten\ and \hs) should be reliable
(Section~\ref{subsec:Contamination}).

In addition to removing bright sources and regions of extended emission, we automatically identified
and removed sources within each field with 0.5--2.0~\kev\ fluxes $\Fsource \ge 1 \times
10^{-14}~\flux$ (cf. $5 \times 10^{-14}~\flux$ in \hs). In general, we obtained the source locations
and fluxes from the 2XMMi DR3 data release of the Second \xmm\ Serendipitous Source
Catalog\footnote{http://xmmssc-www.star.le.ac.uk/Catalogue/2XMMi-DR3/} \citep{watson09}. For this
flux threshold, the catalog is $>$90\%\ complete \citep{watson09}. Note that, although we only used
MOS data in our spectral analysis, the Serendipitous Source Catalog also made use of data from the
pn camera. We excluded the sources identified from the catalog using circles of radius
50\arcsec. Such regions enclose $\sim$90\%\ of each source's flux. In
Section~\ref{subsec:Contamination}, below, we discuss potential contamination from the photons that
spill out of these source exclusion regions.

Approximately 10\%\ of our observations were not included in the Serendipitous Source Catalog.  For
these observations, we ran the source detection ourselves, using the standard
\xmm\ \texttt{edetect\_chain} script.  Following \citet{watson09}, we carried out the source
detection simultaneously in five bands (0.2--0.5, 0.5--1.0, 1.0--2.0, 2.0--4.5, and
4.5--12.0~\kev) using data from the two MOS cameras. For exposures exceeding 5~ks (the minimum
exposure for inclusion in the \hs\ catalog), the MOS cameras can detect sources with
$\Fsource \ge 1 \times 10^{-14}~\flux$ \citep[Figure~3]{watson01}. Again, we excluded the sources
exceeding the flux threshold using circles of radius 50\arcsec.

For each exposure of each observation, we used the \texttt{mos-spectra} script to extract an SXRB
spectrum from the full \xmm\ field of view, minus any excluded sources, and minus any unusable CCDs
(e.g., those in window mode, or those exhibiting the anomalous state described by
\citealt{kuntz08a}). The solid angles of the MOS1 and MOS2 detectors from which the spectra were
extracted are shown in Columns~7 and 9 of Table~\ref{tab:FitResults}, respectively.  We grouped the
SXRB spectra such that each spectral bin contained at least 25 counts. The \texttt{mos-spectra}
script also produced the required response files for each spectrum, namely a redistribution matrix
file (RMF) and an ancillary response file (ARF), using the SAS \texttt{rmfgen} and \texttt{arfgen}
tasks, respectively.

For each SXRB spectrum, we calculated a corresponding quiescent particle background (QPB) spectrum
using the \esas\ \texttt{mos\_back} program. The QPB spectra were constructed from a database of MOS
data obtained with the filter wheel in the closed position, scaled to our observations using data
from the unexposed corner pixels that lie outside the MOS field of view \citep{kuntz08a}. Before we
carried out our spectral analysis, we subtracted from each SXRB spectrum the corresponding QPB
spectrum.

\section{SPECTRAL ANALYSIS METHOD}
\label{sec:Method}

\subsection{Model Description}
\label{subsec:ModelDescription}

We used XSPEC\footnote{http://heasarc.gsfc.nasa.gov/docs/xanadu/xspec/} version 12.7.0k
to carry out the spectral fitting. Our spectral model consisted of components to model the
foreground emission from SWCX (and possibly also from the Local Bubble), the Galactic halo emission
(which is the component that we are interested in here), and the extragalactic background of
unresolved AGN. In addition, the model included components to model the parts of the instrumental
particle background that are not removed by the QPB subtraction. Our model, described below, is the
same as that used in \hskjm, apart from the component used to model the extragalactic background
and, for one sight line, the halo emission model.

\subsubsection{Foreground Emission}
\label{subsubsec:ForegroundModel}

We used a single-temperature ($1T$) \raymondsmith\ model with $T = 1.2 \times 10^6~\K$
\citep{snowden00} to model the foreground emission.  For each sight line, we fixed the normalization
of this component based on the foreground R12 count rates for the five nearest shadows in the
\citet{snowden00} \rosat\ shadow catalog (see \hskjm\ for details).

\subsubsection{Halo Emission}
\label{subsubsec:HaloModel}

We typically used a $1T$ \raymondsmith\ model to model the Galactic halo emission, assuming that the
halo plasma is in collisional ionization equilibrium, and assuming \citet{anders89} solar
abundances.  Although the true halo emission is likely from plasma at a range of temperatures
\citep{yao07a,shelton07,lei09,yao09}, a $1T$ model is generally adequate to characterize the X-ray
emission in the \xmm\ band. We used a \citeauthor{raymond77} model in order to match the code used
to calculate X-ray emission from hydrodynamical models of the halo (\hskjm; D.~B. Henley et~al.,
2013, in preparation). In general, the temperature and emission measure of this component were free
parameters in the fitting. In some cases, typically when the halo emission was faint, XSPEC's
\texttt{error} command was unable to determine the statistical error on the halo temperature. In a
few additional cases, the best-fit temperature was $>$$5 \times 10^6~\K$, but very poorly
constrained. In such cases, we fixed the halo temperature at $2.1 \times 10^6~\K$, and redid the
fit. This temperature was chosen as it was the median halo temperature resulting from our
preliminary analysis of our dataset.

As noted in Section~\ref{subsec:DataReduction}, for nine of our sight lines, we initially found that
fitting with the above $1T$ model yielded a temperature that was unusually high and that was well
constrained, such that $T$ was significantly greater than $4 \times 10^6~\K$. In general, these high
temperatures appeared to be due to excess emission around $\sim$1~\kev, although on about a third of
the sight lines the excess was only slight. As described in Section~\ref{subsec:DataReduction}, we
re-examined these sight lines, using newly created adaptively smoothed, QPB-subtracted images. We
identified and removed additional regions of diffuse emission that could have been contaminating the
spectra. After this modification, only one sight line (number 83) yielded an unusually high halo
temperature ($T \sim \tenMK$), although on some other sight lines, there is still some excess
emission around $\sim$1~\kev\ apparent in the spectra.  For sight line 83 alone, we used a
two-temperature ($2T$) model to model the non-foreground, non-extragalactic-background emission: one
component to model the excess emission around $\sim$1~\kev, and one to model the $\sim$\twoMK\ halo
emission. In the plots and tables that follow, we use the results for the $\sim$\twoMK\ halo
component for this sight line.

\subsubsection{Extragalactic Background}
\label{subsubsec:ExtragalacticModel}

\hskjm\ followed \hsten\ and modeled the extragalactic background as a power-law with a
photon index $\Gamma = 1.46$ and a normalization at 1~\kev\ of
10.5~\pownorm\ \citep{chen97}. However, there is evidence that the extragalactic background steepens
below 1~\kev\ \citep{roberts01}. Furthermore, below 2~\kev, the summed spectrum of the faint sources
that comprise the extragalactic background has $\Gamma = 1.96$ \citep{hasinger93}, compared with
$\Gamma \approx 1.4$ for the 3--10~\kev\ extragalactic background \citep{marshall80}. We therefore
adopted a different model for the extragalactic background here. Specifically, we used the double
broken power-law model from \citet{smith07a}. The first component has a break energy of
$E_\mathrm{b} = 1.2~\kev$, photon indices below and above the break of $\Gamma_1 = 1.54$ and
$\Gamma_2 = 1.4$, respectively, and a normalization of 5.70~\pownorm. The second component has
$E_\mathrm{b} = 1.2~\kev$, with $\Gamma_1 = 1.96$, $\Gamma_2 = 1.4$, and a normalization of
4.90~\pownorm.

We rescaled this model so that its 0.5--2.0~\kev\ surface brightness would be equal to the
integrated surface brightness expected from sources with $\Fsource < 1 \times 10^{-14}~\flux$ (this
is the flux threshold for the automated source removal; see Section~\ref{subsec:DataReduction}).
\citet{hickox06} measured the 0.5--2.0~\kev\ surface brightness of the unresolved extragalactic
background to be $(1.57 \pm 0.41) \times 10^{-12}~\flux\ \pdegsq$, after removing sources with
$\Fsource \ge 2.5 \times 10^{-17}~\flux$. This is the average of the surface brightness measurements
for the \chandra\ Deep Field North (CDF-N) from their Table~3, attenuated by an absorbing column of
$1.5 \times 10^{20}~\pcmsq$ (the value for the CDF-N; \citealt{hickox06}), with the error rescaled
to a 90\%\ confidence interval. The 0.5--2.0~\kev\ surface brightness of sources with $\Fsource =
2.5 \times 10^{-17}$ to $1 \times 10^{-14}~\flux$, meanwhile, is expected to be $2.97 \times
10^{-12}~\flux\ \pdegsq$ (using the source flux distribution from \citealt{moretti03}).  Hence, we
rescaled the \citet{smith07a} extragalactic model such that its observed 0.5--2.0~\kev\ surface
brightness (assuming an absorbing column of $1.5 \times 10^{20}~\pcmsq$, the value for the CDF-N;
\citealt{hickox06}) is $4.54 \times 10^{-12}~\flux\ \pdegsq$. This corresponds to normalizations of
3.59 and 3.09~\pownorm, respectively, for the two components. In Section~\ref{subsec:Systematic},
below, we describe how we estimated the systematic errors associated with our fixing the
normalizations of the extragalactic model components at these nominal values.

The halo and extragalactic components were both subject to absorption. For this purpose we used the
XSPEC \texttt{phabs} model (\citealt{balucinska92}, with an updated He cross-section from
\citealt{yan98}). For each sight line, the absorbing column density was fixed at the \HI\ column
density from the Leiden-Argentine-Bonn (LAB) Survey (\citealt{kalberla05}; values were obtained
using the HEAsoft \texttt{nh} tool).

\subsubsection{Particle Background}

In addition to the above-described SXRB components, our model included components to model any
residual soft proton contamination that remains after the data cleaning described in
Section~\ref{subsec:DataReduction}, and to model the instrumental Al and Si fluorescence lines at
1.49 and 1.74~\kev, respectively. For the former, we used a power-law model that was not folded
through the instrumental response \citep{snowden11}, while for the latter we used two Gaussians.
The parameters for the soft proton power-laws and for the instrumental lines were free parameters
in the fits, and were independent for each MOS exposure. See \hskjm\ for more details.

For each sight line, we fitted the above-described model to the 0.4--5.0~\kev\ MOS1 and MOS2 spectra
simultaneously. In most cases, this involved fitting to the spectra from a single
\xmm\ observation. However, for some sight lines we fitted the model to the spectra from multiple
observations (see Section~\ref{subsec:ObservationSelection}).

\subsection{Systematic Errors}
\label{subsec:Systematic}

In our spectral analysis we fixed both the normalization of the foreground component (estimated
using \rosat\ shadowing data from \citealt{snowden00}) and the normalization of the extragalactic
background (using the surface brightness expected for this background given the source removal flux
threshold; \citealt{moretti03,hickox06}).  We fixed the normalization of the foreground component in
order to avoid having a degeneracy at low energies between the foreground and halo emission, and we
fixed the normalization of the extragalactic background in order to avoid having a degeneracy at
high energies between this component and the power-law used to model the residual soft proton
contamination. Because fixing these model parameters may bias our best-fit halo parameters,
introducing systematic errors to our results, here we estimate the magnitudes of these systematic
errors, using essentially the same method as described in \hskjm\ (although note that \hskjm\ did
not take into account cosmic variance; see below).

To estimate the systematic errors due to our fixing the foreground normalization, we reanalyzed each
sight line with a foreground normalization corresponding to an R12 count rate of 610~\rassrate\ (this
is the median of the values in Column~10 of Table~\ref{tab:FitResults}). We then used the median
differences between the original halo parameters and these new halo parameters to estimate the
systematic errors due to our fixing the foreground normalization, yielding $\pm 0.046 \times
10^6~\K$ and $\pm 0.027$~dex for the halo temperature and emission measure, respectively. We applied
these systematic errors to all sight lines.

Our estimate of the systematic errors due to our fixing the extragalactic normalization is based on
three uncertainties: (1) the uncertainty on the summed surface brightness of sources with $\Fsource
= 2.5 \times 10^{-17}$ to $1 \times 10^{-14}~\flux$ in a given field, due to Poissonian
field-to-field variation of the numbers of such sources (estimable from the \citealt{moretti03}
source flux distribution), (2) the uncertainty on the measured surface brightness of the unresolved
extragalactic background after removing sources with $\Fsource \ge 2.5 \times 10^{-17}~\flux$
\citep{hickox06}, and (3) field-to-field variations in the number of sources comprising the
extragalactic background due to clustering of said sources (cosmic variance). For uncertainty (1),
we used a Monte Carlo simulation to estimate the field-to-field variation in the summed surface
brightness of sources with $\Fsource = 2.5 \times 10^{-17}$ to $1 \times 10^{-14}~\flux$, due to
Poissonian fluctuations in the number of said sources. We estimated this variation to be $\pm 0.22
\times 10^{-12}~\flux\ \pdegsq$ (90\%\ confidence interval) for \xmm-sized fields. For uncertainty
(2), we used the measurement error from \citet{hickox06} quoted in
Section~\ref{subsubsec:ExtragalacticModel}, above: $\pm 0.41 \times 10^{-12}~\flux\ \pdegsq$
(90\%\ confidence interval).

For uncertainty (3), we followed \citet{hickox06}, and used Equation~(45.6) from \citet{peebles80} to
calculate the variance in the number of sources due to source clustering, $\sigma^2_\mathrm{clus}$,
for a field of solid angle $\Omega$:
\begin{equation}
  \frac{\sigma^2_\mathrm{clus}}{N^2} = \int \int w(\theta_{12}) d\Omega_1 d\Omega_2,
  \label{eq:Clustering}
\end{equation}
where $N$ is the expectation value of the number of sources in the field, and $w(\theta_{12})$ is
the two-point angular correlation coefficient. Note that this variance is in addition to that due to
Poissonian fluctuations. \citet{vikhlinin95} found that the clustering of extragalactic X-ray
sources could be described by $w(\theta_{12}) = (\theta_{12}/\theta_0)^{1-\gamma}$, with $\theta_0 =
4\arcsec$ and $\gamma \approx 1.8$. Substituting this into Equation~(\ref{eq:Clustering}) and
integrating over a field of radius 15\arcmin\ (the approximate size of the \xmm\ field of view), the
expected field-to-field variation in the number of extragalactic sources due to clustering is $\pm
14\%$ ($1\sigma$). Assuming that this clustering is independent of source flux, this variation
corresponds to a field-to-field variation in the surface brightness of $\pm 1.03 \times
10^{-12}~\flux\ \pdegsq$ (90\%\ confidence interval).

Combining uncertainties (1) through (3) in quadrature, we estimate that, for \xmm-sized fields, the
0.5--2.0~\kev\ surface brightness of the extragalactic background will typically lie in the range
$(\mbox{3.41--5.66}) \times 10^{-12}~\flux\ \mathrm{deg}^{-2}$ (90\%\ confidence interval). We
therefore reanalyzed each sight line twice, with the extragalactic model rescaled to give
0.5--2.0~\kev\ surface brightnesses of $3.41 \times 10^{-12}$ and $5.66 \times
10^{-12}~\flux\ \mathrm{deg}^{-2}$, respectively (cf. our original extragalactic model had a surface
brightness of $4.54 \times 10^{-12}~\flux\ \mathrm{deg}^{-2}$). During each fit, the surface
brightness of the extragalactic model was fixed at the specified value. We used the differences
between these new results and the original results to estimate the systematic errors (90\%
confidence intervals) due to our fixing the extragalactic normalization. Note that, when we adjust
the normalization of the extragalactic model from its original value, the best-fit soft proton model
changes in response, and the changes in both the components alter the best-fit halo model. Therefore,
this estimated systematic error due to our fixing the extragalactic normalization also takes into
account the uncertainty in the soft proton contamination.

In summary, for each sight line and for each halo model parameter, we estimated the systematic error
due to our fixing the foreground normalization and that due to our fixing the extragalactic
normalization.  We added these errors in quadrature to yield our final estimate of the systematic
error.

\section{SPECTRAL ANALYSIS RESULTS}
\label{sec:Results}

\newcounter{ResultsTable}
\setcounter{ResultsTable}{\value{table}}
\addtocounter{table}{1}

\begin{figure*}
\plottwo{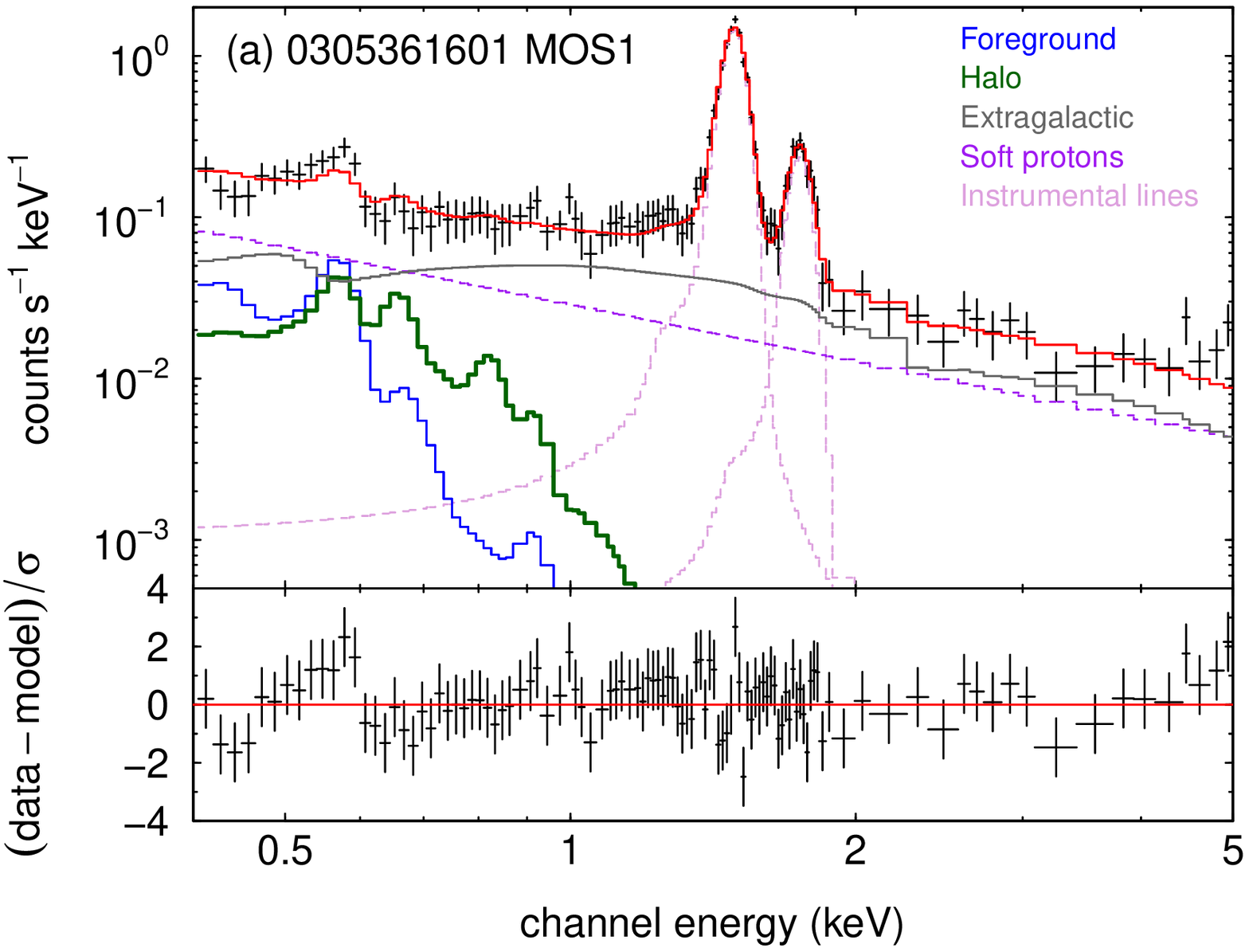}{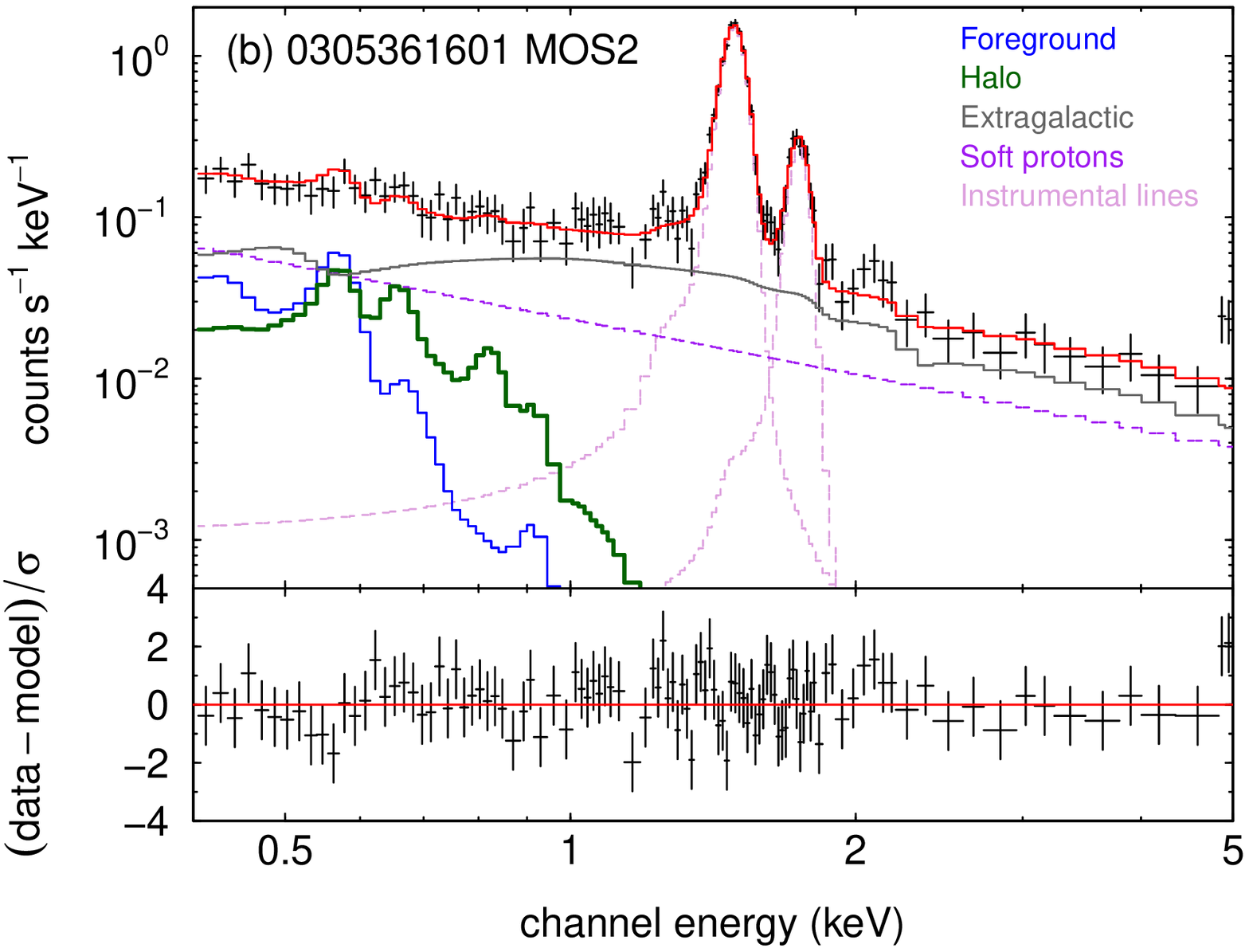}
\plottwo{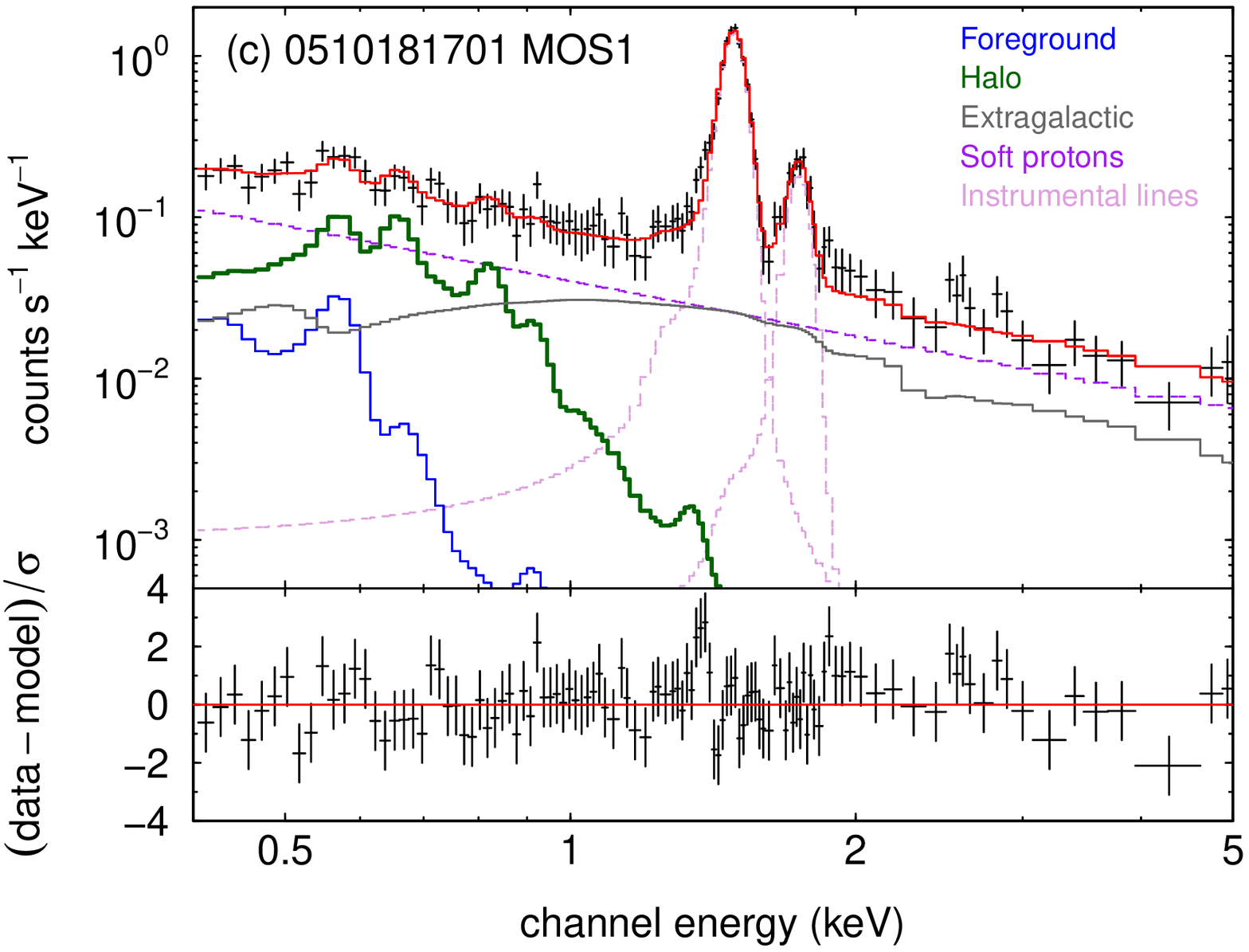}{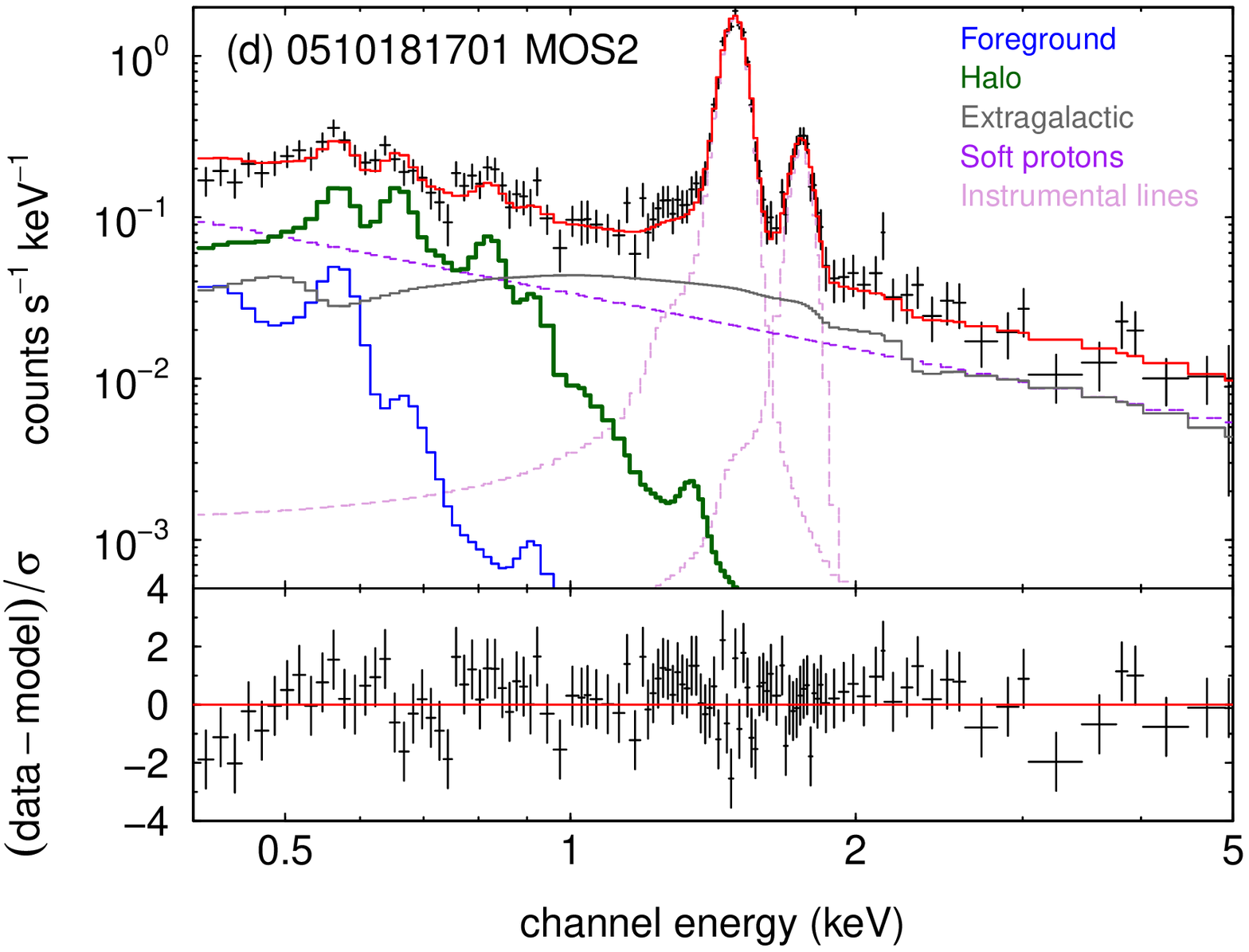}
\caption{\xmm\ MOS1 (left) and MOS2 (right) spectra and best-fit $1T$ halo models for two example
  observations: 0305361601 (sight line 11; top row) and 0510181701 (sight line 97; bottom row). The
  best-fit model is shown in red, and the individual model components are plotted in different
  colors (see key). Note that the two components of the extragalactic background have been summed.
  The components that model parts of the particle background (the soft protons and the instrumental
  lines) are plotted with dashed lines. For the fitting, the spectra were grouped such that there
  were at least 25 counts per bin, prior to subtraction of the QPB. For plotting purposes only, we
  have further grouped the spectra so that each bin has a signal-to-noise ratio of at least 3.
  \label{fig:Spectra1}}
\end{figure*}

\begin{figure}
\plotone{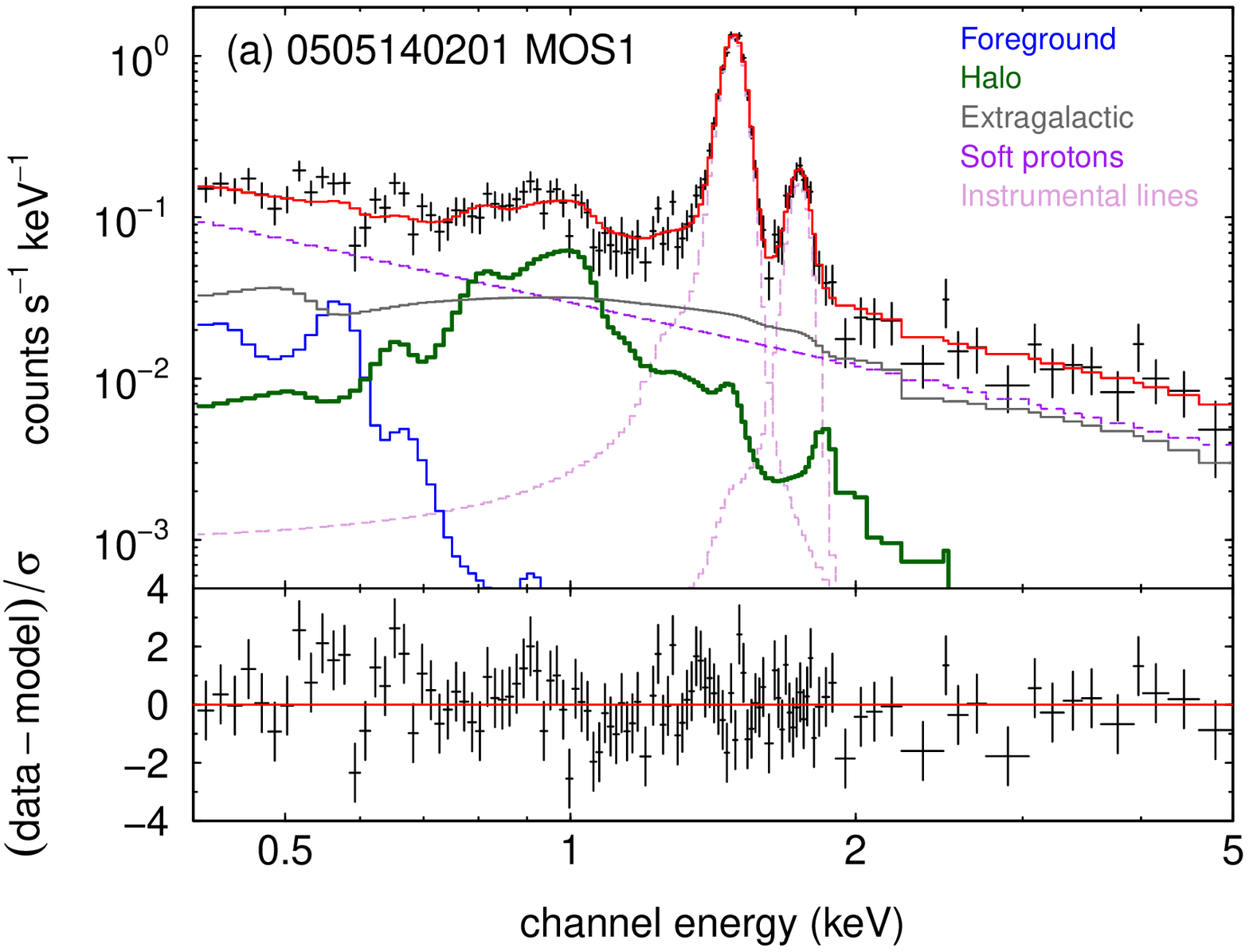}
\plotone{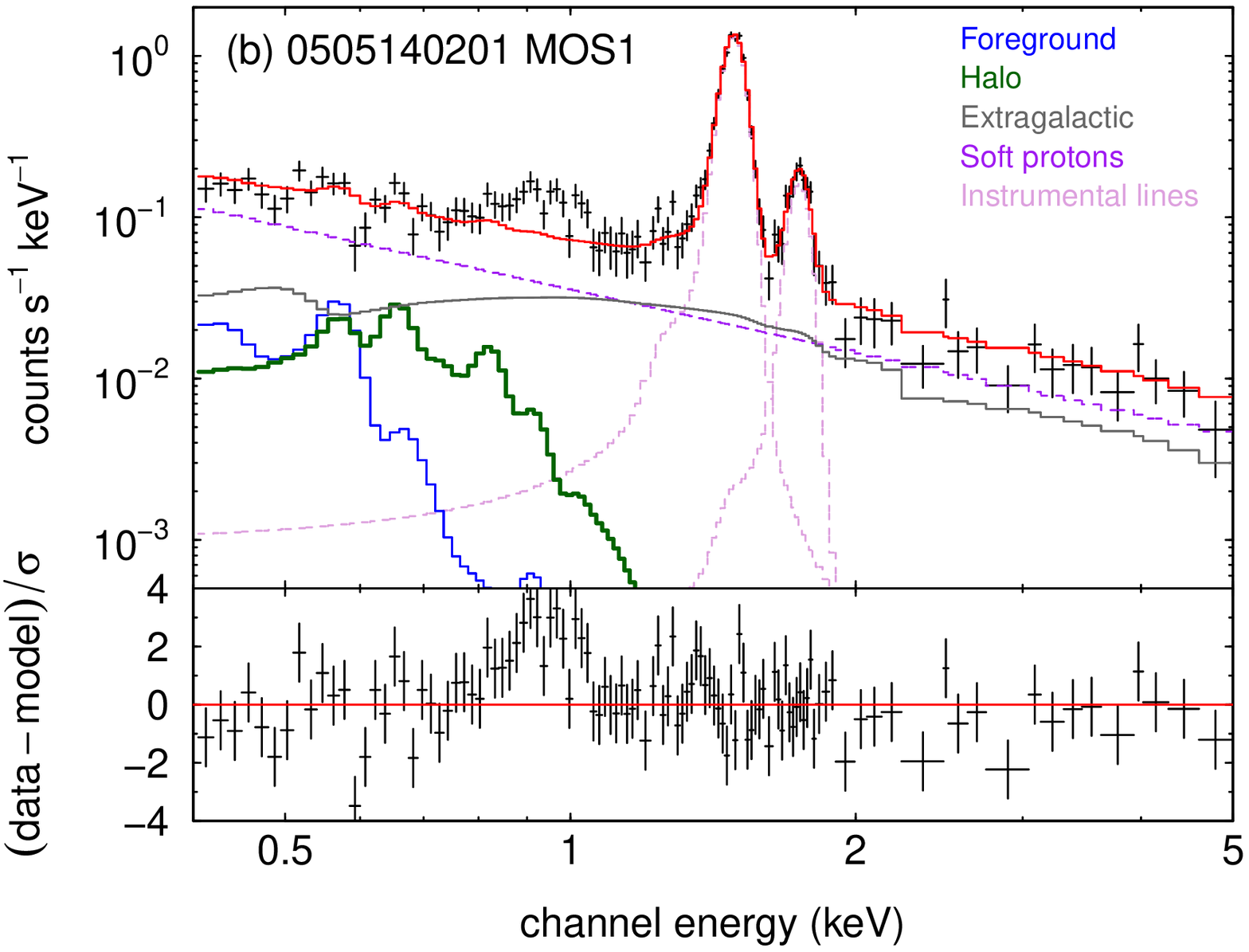}
\plotone{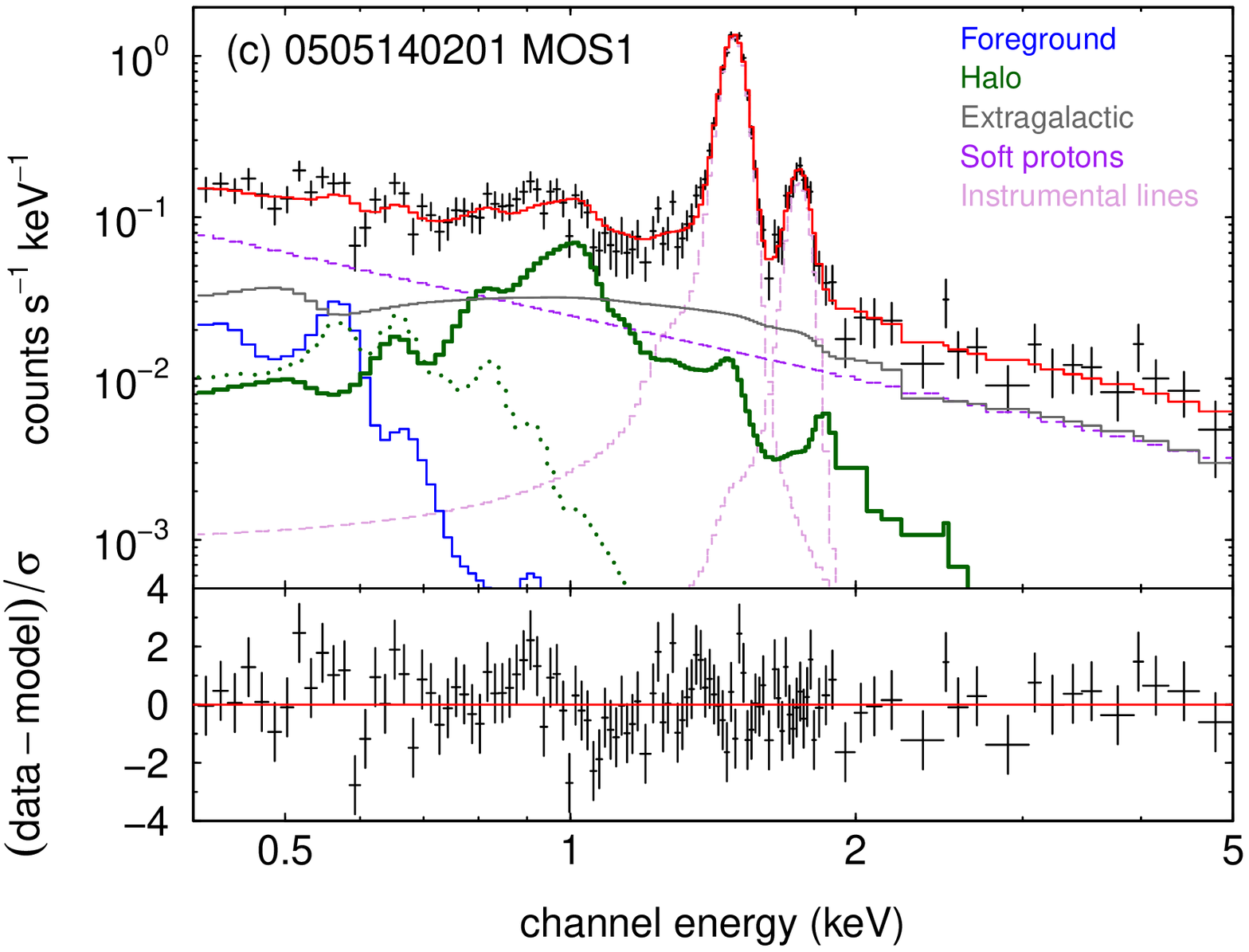}
\caption{\xmm\ MOS1 spectra and best-fit models for observation 0505140201 (from sight line 83).
  The initial fitting on this sight line with a $1T$ halo model yielded an unusually high halo
  temperature ($T \sim \tenMK$; panel (a)). This high temperature is due to excess emission around
  1~\kev; this excess emission is apparent when we fix the halo temperature at $2.5 \times 10^6~\K$
  (panel (b)). For this sight line we used a $2T$ model to model the non-foreground,
  non-extragalactic-background emission (panel (c)). The green dotted line shows the emission from
  the $\sim$\twoMK\ halo, while the green solid line shows the emission from the hotter component, of
  unknown origin. The spectra are grouped as in Figure~\ref{fig:Spectra1}.
  \label{fig:Spectra2}}
\end{figure}

Our spectral modeling inputs and results are shown in Columns~10 through 15 of
Table~\ref{tab:FitResults}. For each sight line with multiple \xmm\ observations, the fit results
are shown against the first listed observation; Columns~10 through 15 are empty for that sight
line's other observations. Column~10 contains the foreground R12 count rate used to fix the
normalization of the foreground component (Section~\ref{subsubsec:ForegroundModel}). Column~11
contains the absorbing \HI\ column density used to attenuate the halo and extragalactic
components. Column~12 contains the best-fit halo temperature, along with the statistical error (90\%
confidence interval for one interesting parameter; \citealt{lampton76}) and the systematic error
(Section~\ref{subsec:Systematic}).  Similarly, Column~13 contains the best-fit halo emission
measure. As noted in Section~\ref{subsec:ObservationSelection}, the results for sight line 103 were
obtained by averaging the results for sight lines 103.1 through 103.27.  Column~14 contains $\chi^2$
and the number of degrees of freedom. Column~15 contains the intrinsic 0.5--2.0~\kev\ surface
brightness of the halo, \Stotal. The \xmm\ spectra and best-fit models for two typical observations
are shown in Figure~\ref{fig:Spectra1}.

We noted in Sections~\ref{subsec:DataReduction} and \ref{subsubsec:HaloModel} that our $1T$ halo
model yielded an unusually high halo temperature for sight line 83 (Figure~\ref{fig:Spectra2}(a)).
This high temperatures appears to be due to excess emission around $\sim$1~\kev\ -- see
Figure~\ref{fig:Spectra2}(b), in which the halo temperature was fixed at $2.5 \times 10^6~\K$. For
this sight line, we used a $2T$ model to model the non-foreground, non-extragalactic-background
emission (Figure~\ref{fig:Spectra2}(c)). The origin of the excess emission around
$\sim$1~\kev\ (which is also apparent on a small number of other sight lines) is uncertain. It is
unlikely to be due to SWCX emission from Fe and Ne. This is because all of our \xmm\ observations
were taken during solar minimum toward high ecliptic latitudes (see
Section~\ref{subsec:ObservationSelection}), and so our observations mostly sample the low-ionization
slow solar wind in which high Fe and Ne ions are not expected \citep{yoshino09}. It is also unlikely
to be due to faint sources below our source removal flux threshold ($\Fsource < 1 \times
10^{-14}~\flux$).  \citet{gupta09a} examined the stacked spectrum of sources with $\Fsource < 1
\times 10^{-14}~\flux$, and found excess emission around 1~\kev.  This excess, which they attributed
to Milky Way stars, could be fitted with a thermal plasma model with $T = 10.7 \times 10^6~\K$.
However, the 0.5--2.0~\kev\ surface brightness of this component is only $0.065 \times
10^{-12}~\flux\ \pdegsq$, which is a factor of $\sim$50 less than the surface brightness of the
hotter component measured on sight line 83. This excess emission may be due to extragalactic diffuse
emission lying in the field of view, although we have attempted to minimize such emission. However,
the presence of such emission on sight line 83 (and potentially on other sight lines) appears not to
be contaminating our measurements of the $\sim$\twoMK\ halo plasma (see
Section~\ref{subsec:Contamination}).

\begin{figure*}
\plotone{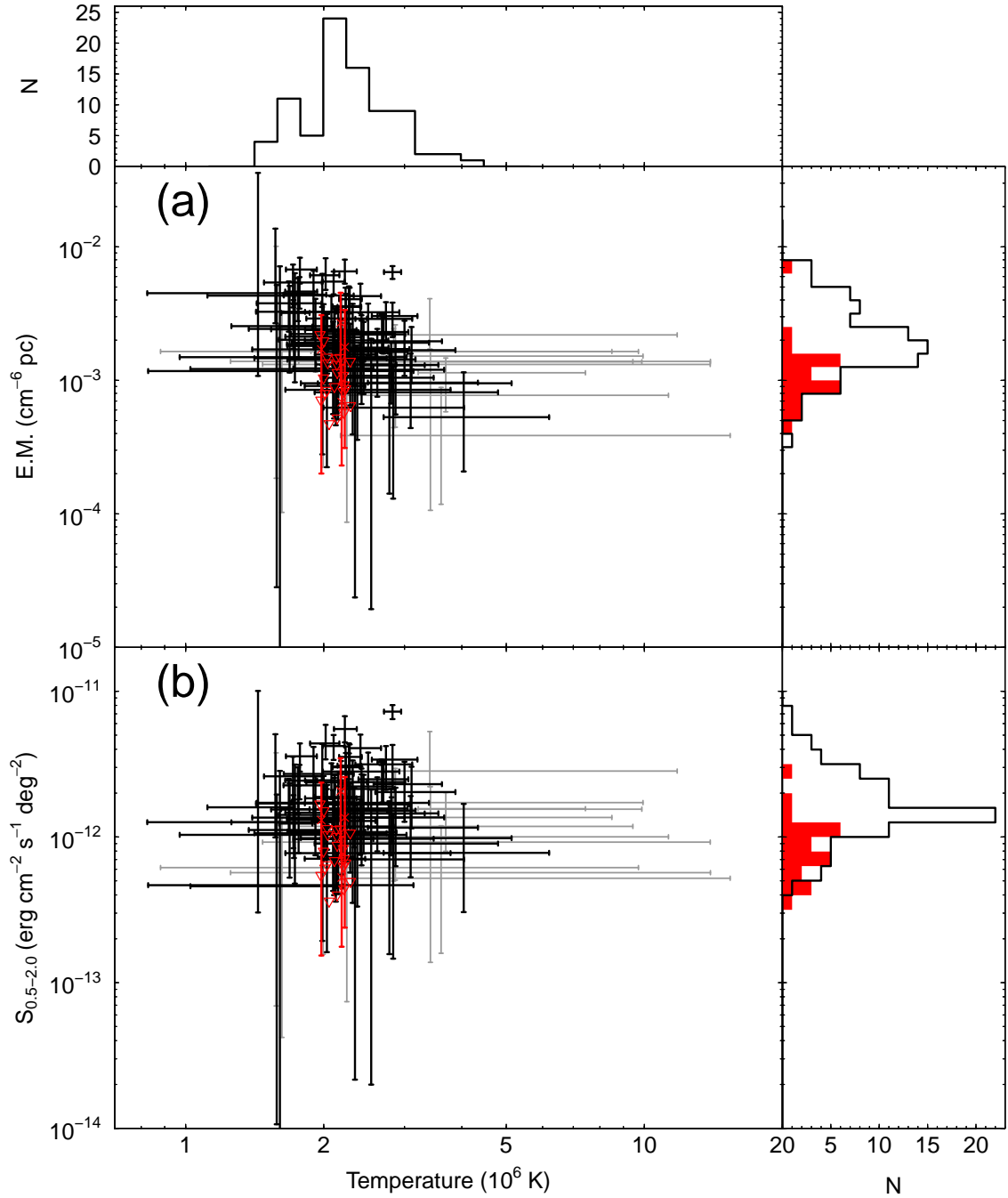}
\caption{(a) Halo emission measure and (b) intrinsic 0.5--2.0~\kev\ halo surface brightness against
  halo temperature, from our spectral modeling.
  \textit{Black}: The temperature was free to vary, and is well constrained.
  \textit{Gray}: The temperature was free to vary, but is poorly constrained (combined statistical
  and systematic confidence interval spans more than $4 \times 10^6~\K$),
  \textit{Red}: The temperature was fixed at $T = 2.1 \times 10^6~\K$ (see
  Section~\ref{subsubsec:HaloModel}).  The red triangles indicate upper limits on the emission
  measures and surface brightnesses. Note that, to avoid clutter, the red data points have been
  randomly displaced by small amounts in the horizontal direction from $T = 2.1 \times 10^6~\K$.
  \textit{Top panel}: Histogram of halo temperatures. The sight lines on which the temperature was fixed
  have been omitted.
  \textit{Side panels}: Histograms of halo emission measures (\textit{upper panel}) and intrinsic
  surface brightnesses (\textit{lower panel}). \textit{Black}: Detections; \textit{red}: upper limits.
  \label{fig:EMvT}}
\end{figure*}

\begin{figure*}
\plottwo{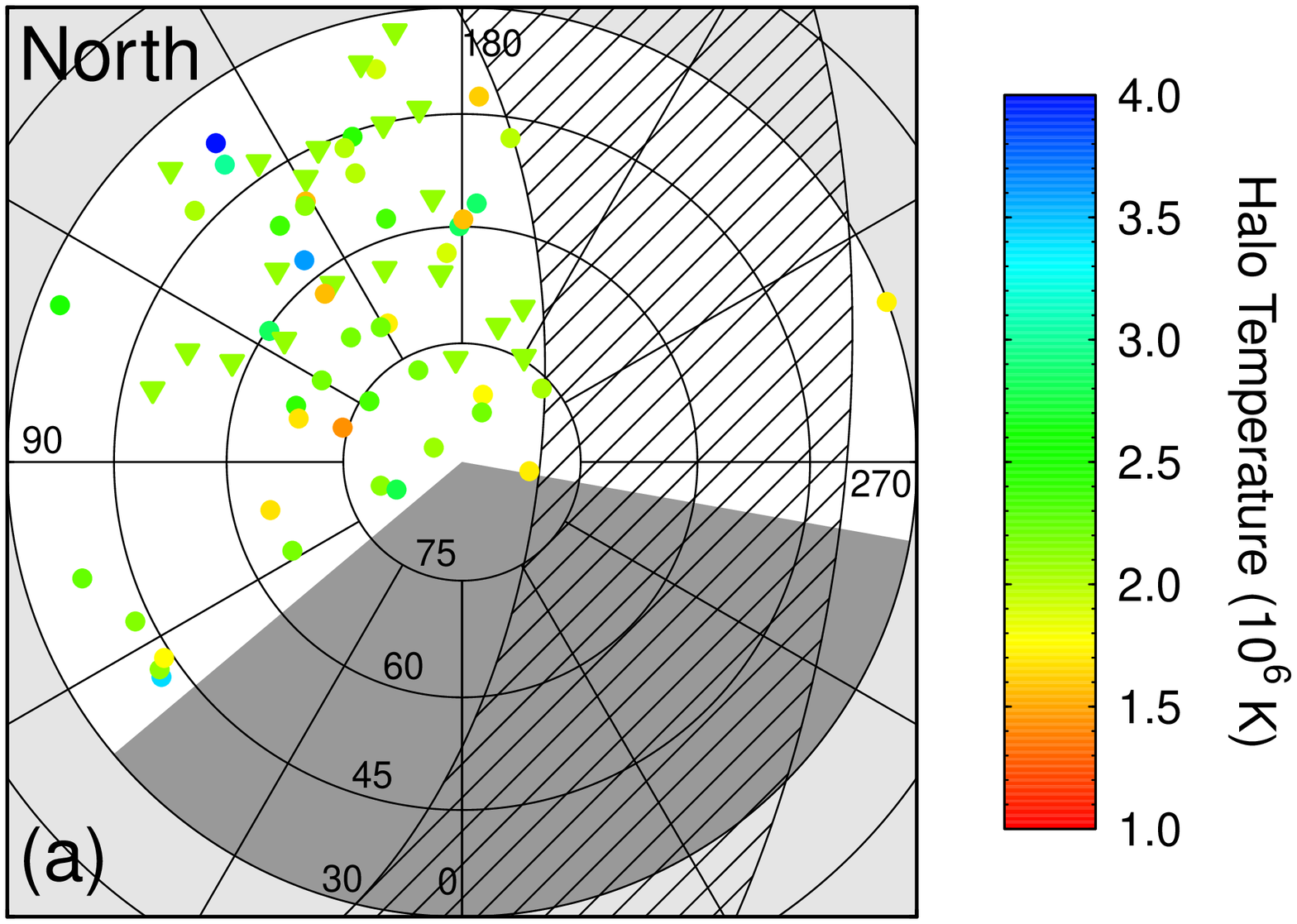}{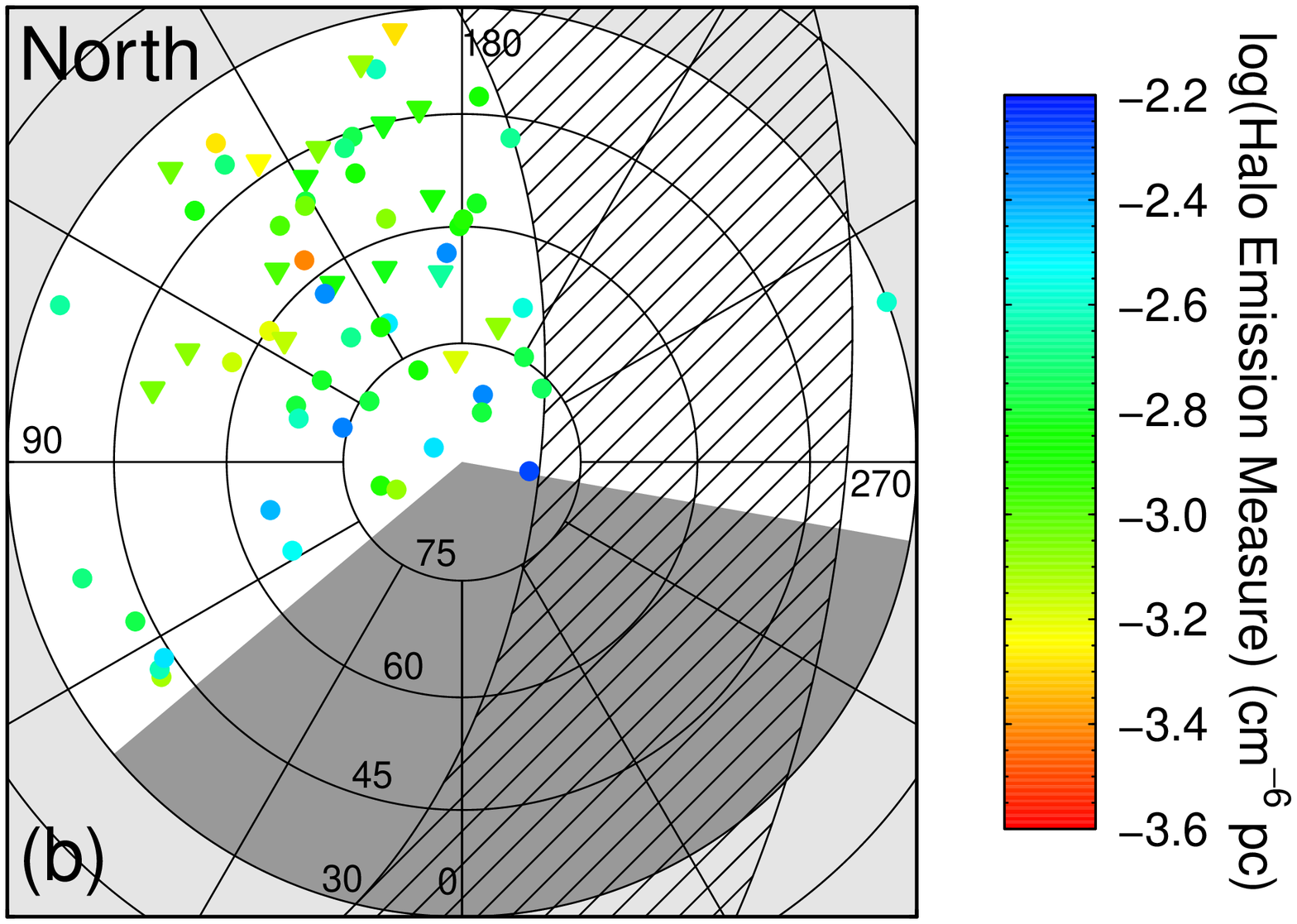}
\plottwo{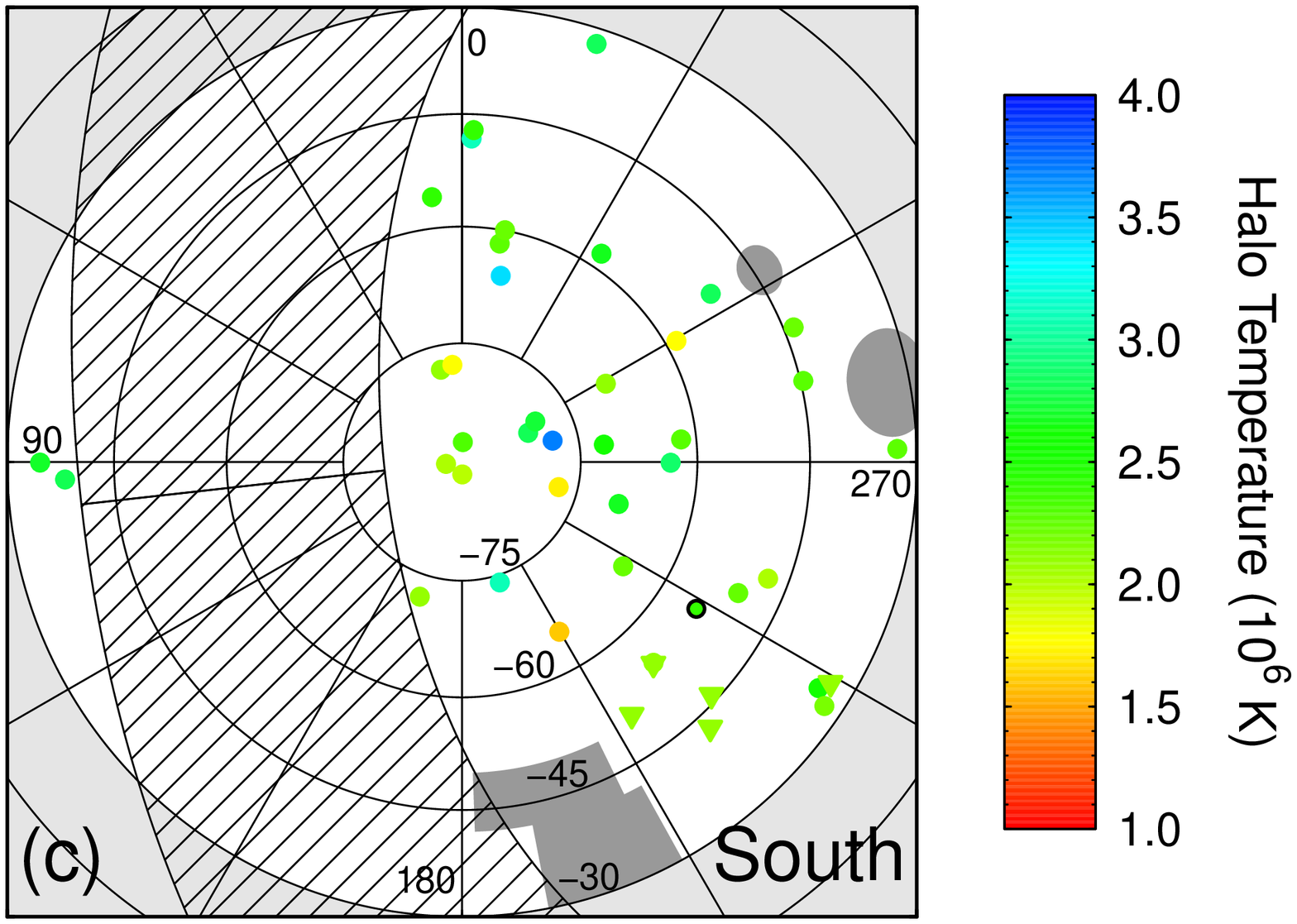}{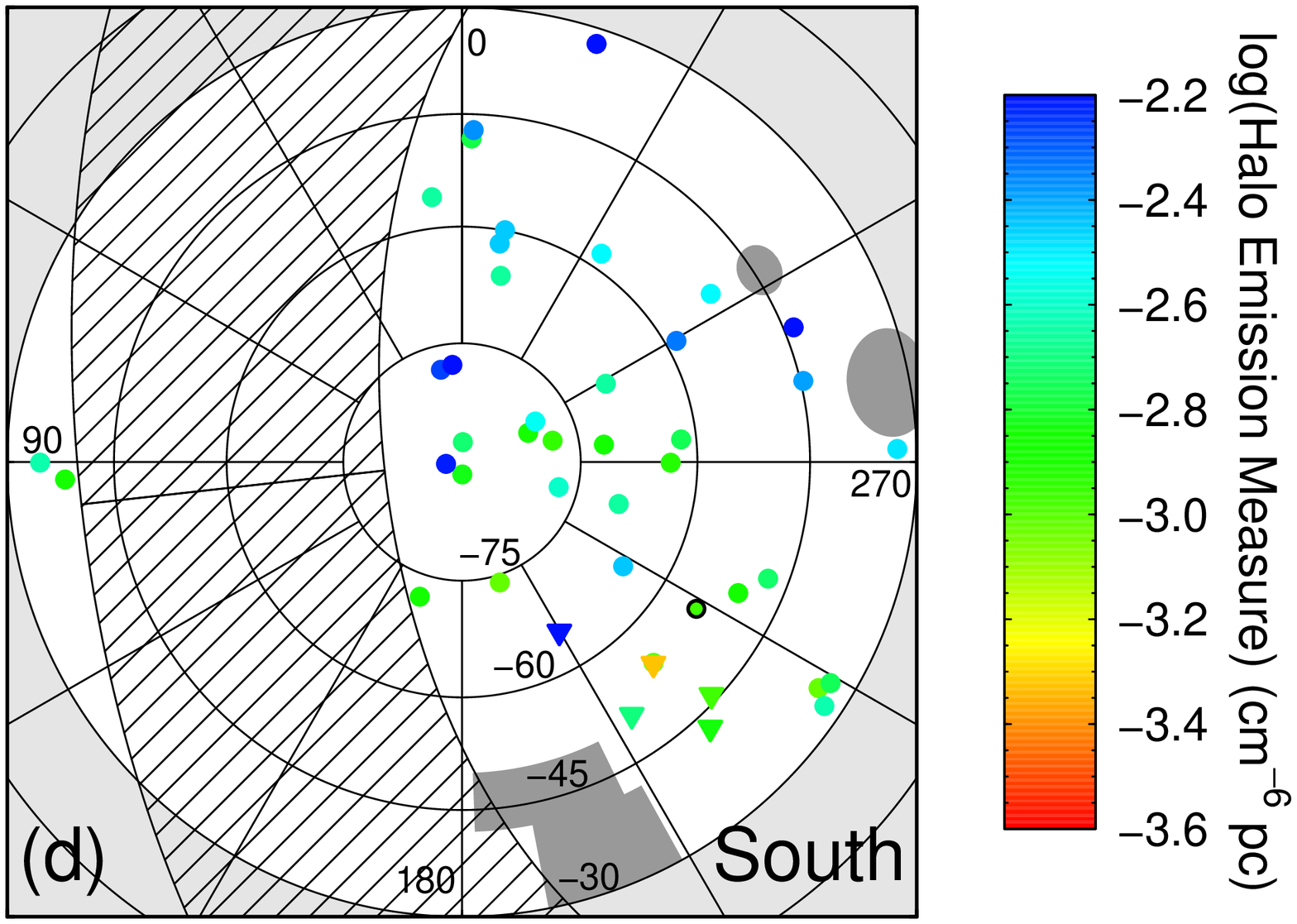}
\caption{Zenith equal-area maps showing the (a) temperatures and (b) emission measures of the halo
  in the northern Galactic hemisphere obtained from our spectral modeling. $l=0\degr$ is toward the
  bottom of the figures, and $l$ increases clockwise. The light gray area indicates the region with
  $|b| \le 30\degr$, the dark gray area indicates the exclusion region around the Sco-Cen
  superbubble, and the hatched area indicates the region with $|\beta| \le 20 \degr$.  Each of these
  regions are excluded in this analysis.  In the temperature map, the triangles indicate that the
  temperature was fixed at $2.1 \times 10^6~\K$.  In the emission measure map, the triangles
  indicate upper limits on the halo emission measure.  Panels (c) and (d) show the corresponding
  results for the southern Galactic hemisphere. Note that $l=0\degr$ is toward the top of the
  figures, and $l$ increases counterclockwise. The dark gray areas indicate the exclusion regions
  around the Eridanus Enhancement, the Large Magellanic Cloud, and the Small Magellanic Cloud, in
  order of increasing longitude. Sight line 83, which was analyzed with a $2T$ model
  (see Section~\ref{subsubsec:HaloModel}) is outlined in black; we have plotted the results for the
  cooler ($T \sim \twoMK$) component for this sight line
  \label{fig:Maps}}
\end{figure*}

The halo emission measures and intrinsic surface brightnesses are plotted against the halo
temperatures in Figure~\ref{fig:EMvT}, with marginal histograms showing the distributions of these
quantities. We define halo emission as having been detected on a sight line if the combined
statistical and systematic confidence interval on the emission measure does not include zero.
Overall, we detected emission from plasma with $T \sim \twoMK$ on 87 of our 110 sight lines (79\%).
For the vast majority of the sight lines with such detections (83/87), we did not have to fix the
halo temperature at $2.1 \times 10^6~\K$. For sight lines with detections, the temperature of the
halo is typically $(\mbox{2.0--2.6}) \times 10^6~\K$ (Table~\ref{tab:Quartiles}, row~1).  The
corresponding emission measures span an order of magnitude, lying mostly in the range
$\sim$$(\mbox{0.8--5}) \times 10^{-3}~\emismeas$, with lower and upper quartiles of $1.4 \times
10^{-3}$ and $3.0 \times 10^{-3}~\emismeas$, respectively (Table~\ref{tab:Quartiles}, row~1). The
intrinsic 0.5--2.0~\kev\ surface brightnesses lie mostly in the range $\sim$$(\mbox{0.6--4}) \times
10^{-12}~\flux\ \pdegsq$, with lower and upper quartiles of $1.1 \times 10^{-12}$ and $2.3 \times
10^{-12}~\flux\ \pdegsq$, respectively (Table~\ref{tab:Quartiles}, row~1).

For 22 of the 23 sight lines on which we did not detect halo emission, we had to fix the halo
temperature at $2.1 \times 10^6~\K$. For the other sight line (number 72), we were able to constrain the halo
temperature because the statistical error on the emission measure alone does not include zero
(although the combined statistical and systematic error does include zero).  Among the sight lines
that yielded non-detections, the lower and upper quartiles of the upper limits on the emission
measures are $0.8 \times 10^{-3}$ and $1.4 \times 10^{-3}~\emismeas$, respectively, while the lower
and upper quartiles of the upper limits on the intrinsic 0.5--2.0~\kev\ surface brightnesses are
$0.6 \times 10^{-12}$ and $1.1 \times 10^{-12}~\flux\ \pdegsq$, respectively
(Table~\ref{tab:Quartiles}, row~2).

Figure~\ref{fig:Maps} shows maps of the measured halo temperatures and emission measures. From a
visual inspection of Figures~\ref{fig:Maps}(a) and (c), it appears that the halo temperature is in
general rather uniform. Figures~\ref{fig:Maps}(b) and (d), meanwhile, show that there is
considerable variation in the emission measure of the halo plasma. In the northern hemisphere, no
clear trends are apparent from Figure~\ref{fig:Maps}(b) (although see
Section~\ref{subsec:LBVariation}, below).  From Figure~\ref{fig:Maps}(d), it appears that the halo
emission measure in the south tends to increase from the outer Galaxy ($l = 180\degr$) to the inner
Galaxy ($l = 0\degr$), a trend which we will confirm in the following section.

\input{tab2}

\subsection{Variation with Galactic Longitude and Latitude}
\label{subsec:LBVariation}

\begin{figure*}
\plottwo{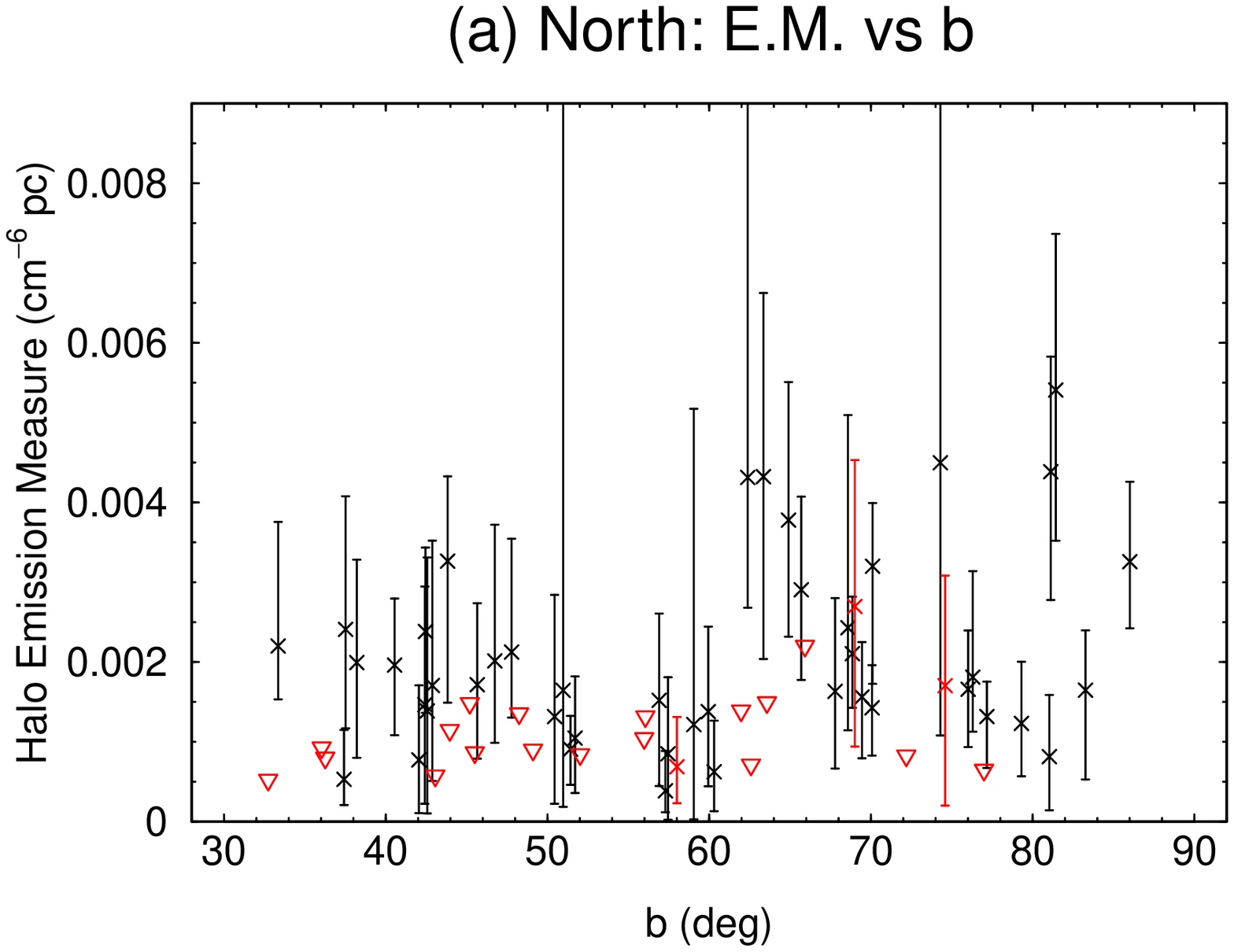}{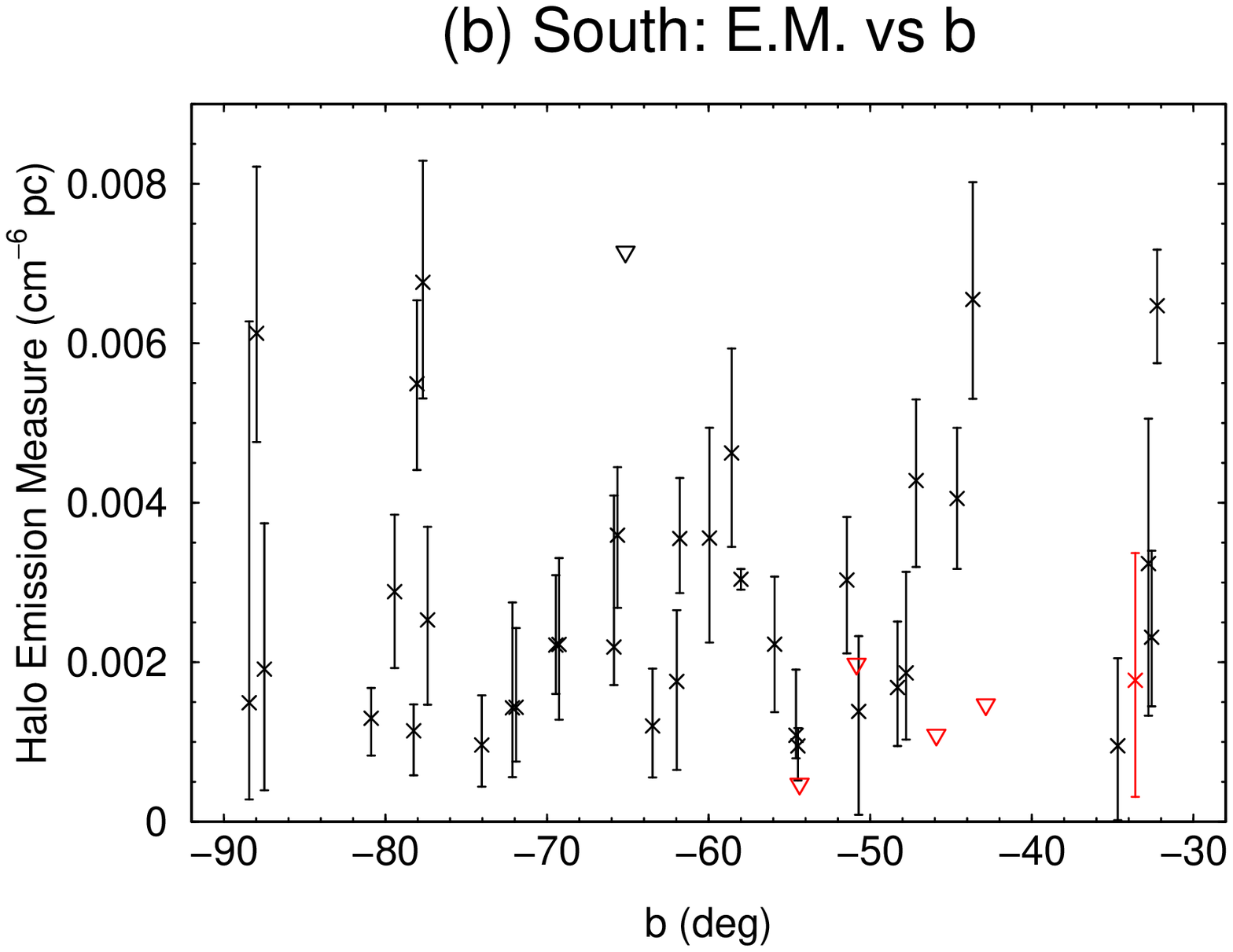}
\plottwo{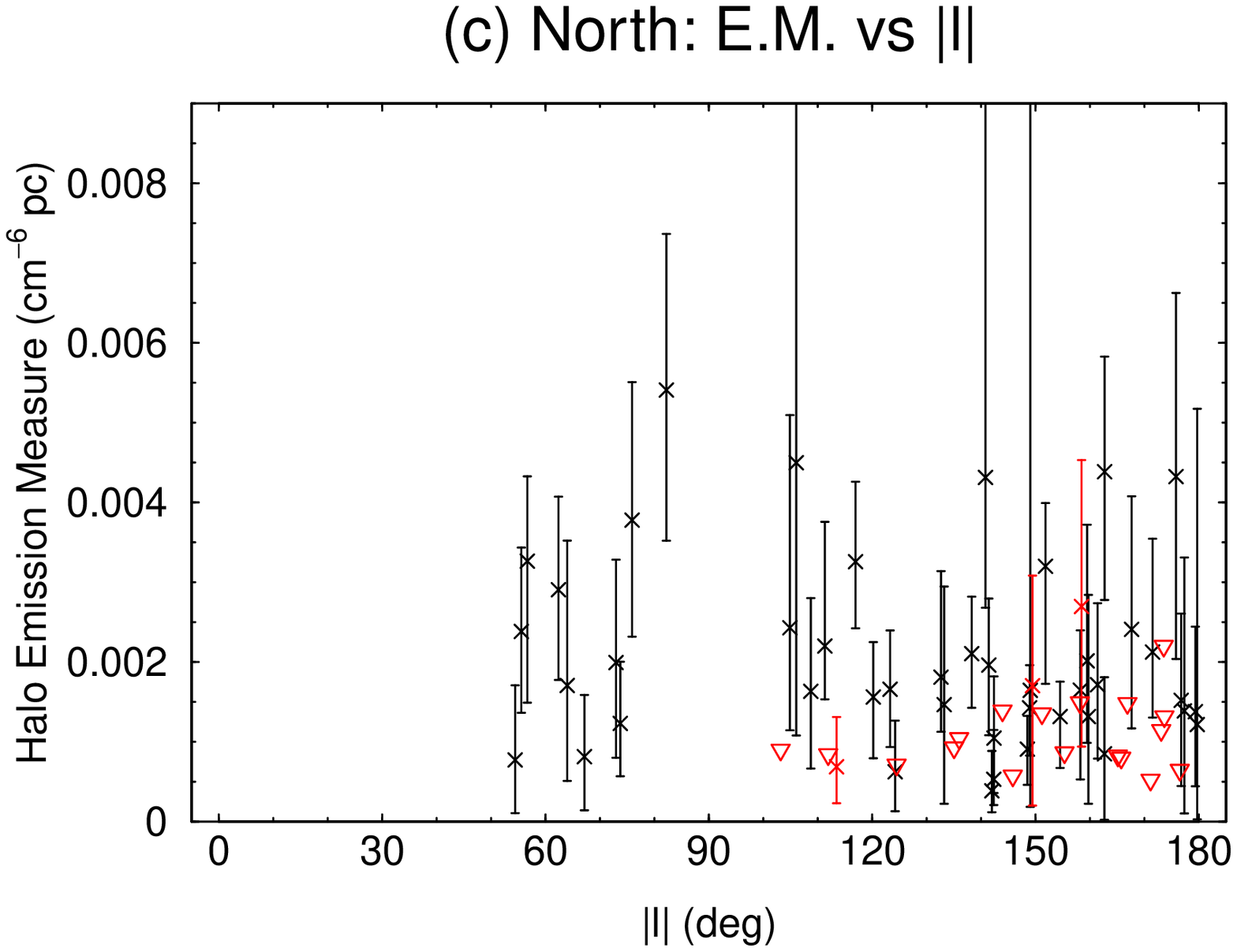}{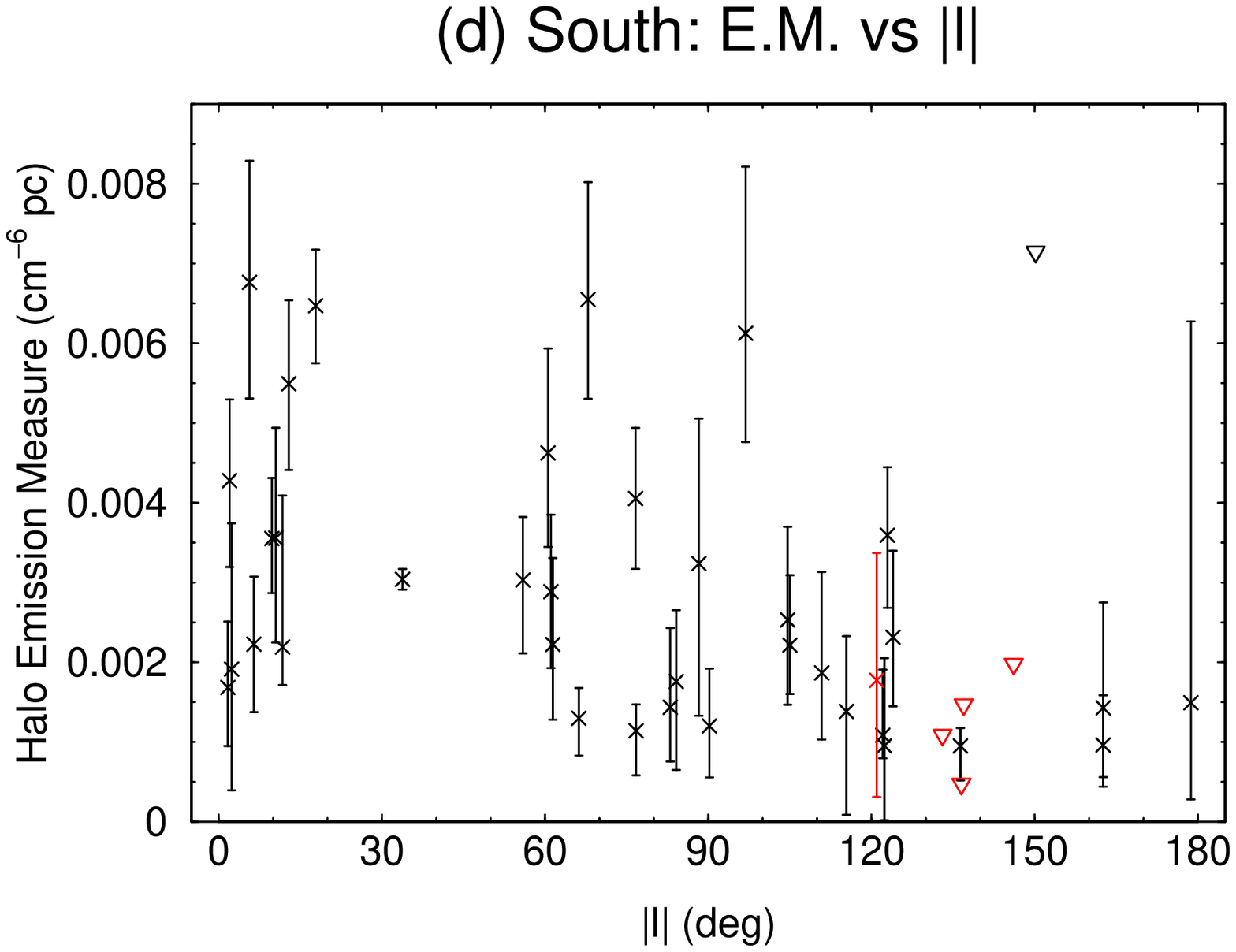}
\caption{Halo emission measure against
  (a) $b$ for the northern Galactic hemisphere,
  (b) $b$ for the Southern Galactic hemisphere,
  (c) $|l|$ for the northern Galactic hemisphere (see Equation~(\ref{eq:lprime})), and
  (d) $|l|$ for the southern Galactic hemisphere.
  Panels~(a) and (d) show the only two examples of statistically significant correlations
  in Table~\ref{tab:CorrelationCoefficients}; the other two panels are shown for
  comparison. Detections are shown with crosses and error bars, upper limits are shown with
  triangles. The red data points indicate emission measures from sight lines for which the
  temperature was fixed at $2.1 \times 10^6~\K$.
  \label{fig:Correlation}}
\end{figure*}

\input{tab3}

Table~\ref{tab:CorrelationCoefficients} shows the correlation coefficients (Kendall's $\tau$; e.g.,
\citealt{press92}) for the halo temperature and emission measure against the absolute values of
Galactic longitude and latitude, $|l|$ and $|b|$, respectively. We define the absolute value of
longitude as
\begin{equation}
  |l| = \left\{
  \begin{array}{ll}
    l            & \mbox{if $l < 180\degr$;} \\
    360\degr - l & \mbox{otherwise;}
  \end{array}
  \right.
  \label{eq:lprime}
\end{equation}
i.e., in both the western and eastern Galactic hemispheres, $|l|$ increases from 0\degr\ toward the
Galactic Center to 180\degr\ toward the Galactic Anticenter. When calculating the correlation
coefficients for the halo temperature against $|l|$ or $|b|$, we omitted the sight lines for which
we had to fix the temperature at $2.1 \times 10^6~\K$.  When calculating the correlation
coefficients for the halo emission measure against $|l|$ or $|b|$, we did not omit such sight
lines. We used the best-fit emission measures for all sight lines, whether or not a given sight line
yielded a detection or an upper limit. Note from Figure~\ref{fig:Maps} that in each hemisphere there
are one or two sight lines that are isolated from the majority of the sight lines in that
hemisphere: in the north there is a single sight line at $(l,b) \approx (250\degr,+30\degr)$, and in
the south there is a pair of sight lines near $(l,b) = (90\degr,-35\degr)$. These sight lines were
excluded from the correlation coefficient calculations.

In only two cases is the correlation statistically significant at the 5\% level. The halo emission
measure is positively correlated with $|b|$ in the northern hemisphere (i.e., the emission measure
tends to increase from low latitudes to the pole; see Figure~\ref{fig:Correlation}(a)).
There is no correlation between emission measure and $|b|$ in the southern hemisphere (see
Figure~\ref{fig:Correlation}(b)). However, the emission measure is negatively correlated
with $|l|$ in the southern hemisphere (i.e., the emission measure tends to increase from the outer
Galaxy to the inner Galaxy, as noted in the previous section; see
Figure~\ref{fig:Correlation}(d)).

The correlation between emission measure and $|b|$ in the north (Figure~\ref{fig:Correlation}(a))
may in part be due to the fact that there are more upper limits below 60\degr\ than above
60\degr. For such sight lines, we used the best-fit emission measures (which are zero for several
sight lines). If, instead, we use the upper limits on the emission measures in the correlation
coefficient calculation, the correlation is still significant, albeit with a higher \pvalue\ ($\tau
= 0.17$, $\pvalue = 0.044$). However, if we omit the sight lines that yield upper limits, the
correlation is not statistically significant ($\tau = 0.11$, $\pvalue = 0.27$).

In contrast, the correlation between emission measure and $|l|$ in the south
(Figure~\ref{fig:Correlation}(d)) remains if we use the upper limits for sight lines with
non-detections ($\tau = -0.34$, $\pvalue = 1.6 \times 10^{-3}$) or if we omit the non-detections
altogether ($\tau = -0.36$, $\pvalue = 1.7 \times 10^{-3}$).  The exclusion of the region around the
Sco-Cen superbubble may limit our ability to detect a similar trend in the northern hemisphere, as
observations with smaller values of $|l|$ are excluded from our dataset (see
Figure~\ref{fig:Correlation}(c)). However, it should be noted that if we exclude a similar region
toward the inner Galaxy in the southern hemisphere ($|l| \le 50\degr$), the correlation between
emission measure and $|l|$ remains significant ($\tau = -0.46$, $\pvalue = 1.5 \times
10^{-5}$). This suggests that, rather than the lack of a significant observed correlation being an
artifact of the exclusion of the region around the Sco-Cen superbubble, the emission measure does
not vary systematically with $|l|$ in the northern hemisphere is it does in the south.

\subsection{Differences between the Galactic Hemispheres}
\label{subsec:HemisphereDifferences}

In the previous subsection we found that the two Galactic hemispheres are different when it comes to
correlations of the halo emission measure with $|l|$ and $|b|$. In this subsection we describe
evidence of other differences between the hemispheres.

There is some evidence (not apparent in the maps in Figure~\ref{fig:Maps}) that the halo tends to be
slightly hotter in the southern hemisphere than in northern hemisphere (see rows~4 and 3 of
Table~\ref{tab:Quartiles}, respectively). A Mann-Whitney $U$ test \citep[e.g.,][]{barlow89,wall03}
indicates that the difference in the median temperatures from the two hemispheres is statistically
significant at the 1\% level ($U = 561$, $\pvalue = 0.0073$ (two-sided)). We pointed out in
Section~\ref{subsec:LBVariation}, above, that the region toward the inner Galaxy is excluded in the
northern hemisphere but not in the southern hemisphere, which could affect the comparison of the
hemispheres. If we exclude the region with $|l| \le 50\degr$ in the south
(Table~\ref{tab:Quartiles}, row~5), we find that the difference in the median temperatures from the
two hemispheres is still statistically significant, but now only at the 5\% level ($U = 417$,
$\pvalue = 0.027$ (two-sided)). However, it should be noted that the difference is less than the
typical error on the temperature ($\sim \pm 0.4 \times 10^6~\K$).

There are more non-detections of X-ray emission from the halo in the northern
hemisphere than in the south: there are non-detections on 18 out of 66 sight lines (27\%) in the
north compared with 5 out of 44 sight lines (11\%) in the south. However, Fisher's exact test
\citep[e.g.,][]{wall03} indicates that this difference between the hemispheres is not statistically
significant ($\pvalue = 0.056$ for a two-sided test). A related fact is that the halo emission
measure tends to be lower in the northern hemisphere than in the south (see rows~3 and 4 of
Table~\ref{tab:Quartiles}, respectively). In this case, the difference is significant: the median
emission measure in the north is significantly lower than that in the south at the 1\% level
(Mann-Whitney $U = 971$, $\pvalue = 0.0034$ (two-sided)). However, this difference between the
hemispheres may be due to the exclusion of the region toward the inner Galaxy in the northern
hemisphere (note that the emission measure increases toward the inner Galaxy in the southern
hemisphere; see Section~\ref{subsec:LBVariation}, above).  If we exclude the region with $|l| \le
50\degr$ in the south, the median emission measure in the south (Table~\ref{tab:Quartiles}, row~5)
is no longer significantly different from that in the north (Table~\ref{tab:Quartiles}, row~3; $U =
868$, $\pvalue = 0.10$).

In summary, there is some evidence that the halo temperature and emission measure tend to be
somewhat higher in the south than in the north. However, these differences may in part be due to the
fact that we do not have data from equivalent regions of the halo in the two hemispheres, as the
region toward the inner Galaxy is excluded in the north (because of the presence of the Sco-Cen
superbubble).

\section{DISCUSSION}
\label{sec:Discussion}

In the following subsections, we compare our measurements with those from previous studies, we
discuss the effect of our assumed foreground model on our halo measurements, and we consider sources
of contamination that could be affecting our halo measurements
(Sections~\ref{subsec:ComparisonWithPreviousStudies} through \ref{subsec:Contamination},
respectively). We conclude that contamination and our choice of foreground model are, in general,
not adversely affecting our halo measurements. Then, in Section~\ref{subsec:Morphology}, we discuss
the morphology of the hot halo. Finally, in Section~\ref{subsec:Origins}, we comment on the
implications of our measurements for the origin of the hot halo (deferring a more detailed study of
this issue to a follow-up paper; D.~B. Henley et~al., 2013, in preparation).

\subsection{Comparison with Previous Studies}
\label{subsec:ComparisonWithPreviousStudies}

As noted in Section~\ref{subsec:ObservationSelection}, our sample of \xmm\ observations includes 20
that were analyzed in the \hskjm. The halo temperatures that we have measured for these sight lines
are generally in good agreement with those measured in \hskjm, and there is no systematic difference
in the halo temperatures (although it should be noted that for five of these sight lines we had to
fix the halo temperature at $2.1 \times 10^6~\K$ for the current analysis).

The current analysis does, however, yield systematically lower emission measures and surface
brightnesses than \hskjm, typically by about a third. Although we use a lower source removal
flux threshold than in \hskjm\ ($1 \times 10^{-14}$ versus $5 \times 10^{-14}~\flux$), this appears
not to be directly responsible for the difference (i.e, sources with $\Fsource = 1 \times 10^{-14}$
to $5 \times 10^{-14}~\flux$ were not contaminating the \hskjm\ halo measurements). Instead, the
difference is most likely due to our using a lower normalization for the extragalactic background
(the extragalactic normalization used in \hsten\ and \hskjm\ may have been too large,
given the source removal threshold). This lower extragalactic normalization results in a higher
normalization for the soft proton model, in order to match the observed count rate above
$\sim$2~\kev. Because of these two components' different spectral shapes, these changes result in
more counts below $\sim$1~\kev\ being attributed to the combination of these two components, and
hence in fewer counts being attributed to the halo emission (we noted a similar effect in \hs\ when
we compared our oxygen intensity measurements with those from \hsten). Therefore, the
presence of the soft proton contamination in the \xmm\ spectra, which requires us to fix the
normalization of the extragalactic background, potentially introduces some uncertainty in the
normalization of the halo X-ray emission.

\begin{figure}
\plotone{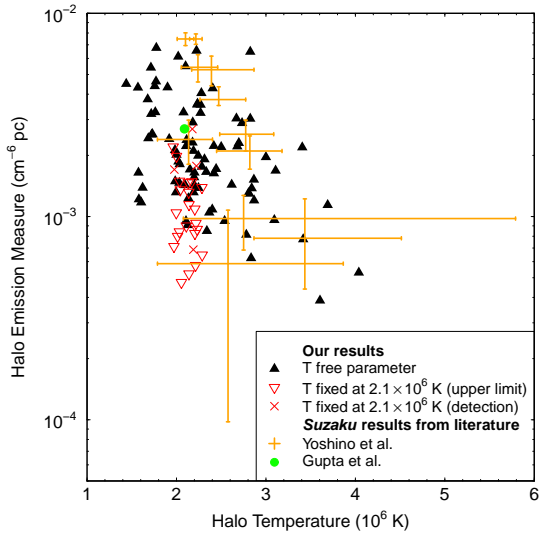}
\caption{Comparison of our halo results with those obtained from \suzaku\ SXRB spectra.
  Our results are generally plotted with black triangles, with the error bars omitted. If the
  temperature was fixed at $2.1 \times 10^6~\K$, the results are plotted with red triangles and red
  crosses, for upper limits and detections, respectively. As in Figure~\ref{fig:EMvT}, the red data
  points have been staggered in the horizontal direction from $T = 2.1 \times 10^6~\K$, in order to
  reduce clutter.
  The results from the \citet{yoshino09} \suzaku\ study are plotted with orange crosses (from their
  Table~6). We only show \citeauthor{yoshino09}'s measurements of the $\sim$\twoMK\ halo; we do not
  plot the results for the hotter halo component from their Table~7. We have omitted the
  low-latitude LL10 sight line, and sight line LH-2, on which halo emission is not detected.
  The light green circle near $T = 2 \times 10^6~\K$, $\mbox{emission measure} = 3 \times
  10^{-3}~\emismeas$ shows the result from the \suzaku\ shadowing observation of MBM~20
  \citep{gupta09b}.
  \label{fig:CompareOtherObs}}
\end{figure}

In Figure~\ref{fig:CompareOtherObs}, we compare our \xmm\ halo measurements with \suzaku\ halo
measurements from the literature. We have plotted our temperature and emission measure measurements,
along with the values measured by \citet{yoshino09} from 11 high-latitude \suzaku\ observations (the
largest study of the SXRB with \suzaku\ to date). We also plot the temperature and emission measure
found by \citet{gupta09b} from their \suzaku\ shadowing study of MBM~20 -- we included this data
point so that we could compare different methods for determining the foreground emission (see
Section~\ref{subsec:ForegroundModel}, below). Note that the observations used by \citet{smith07a}
and \citet{lei09} in their \suzaku\ shadowing analyses are included in the \citet{yoshino09}
dataset, and so we do not include \citeauthor{smith07a}'s or \citeauthor{lei09}'s results in
Figure~\ref{fig:CompareOtherObs}.

The ranges of temperatures and emission measures measured by
\citet{yoshino09} and \citet{gupta09b} are generally similar to ours, although our dataset includes
several sight lines with $T < 2 \times 10^6~\K$, which the \suzaku\ dataset does not. While the
median temperature ($2.43 \times 10^6~\K$) and median emission measure ($2.62 \times
10^{-3}~\emismeas$) from the combined \citeauthor{yoshino09}\ and \citeauthor{gupta09b}\ dataset are
both somewhat higher than our median detected values ($2.22 \times 10^6~\K$ and $1.91 \times
10^{-3}~\emismeas$, respectively; Table~\ref{tab:Quartiles}, row~1), Mann-Whitney $U$ tests indicate
that these differences are not statistically significant ($U = 638$ and 654, with two-sided
$p$~values of 0.14 and 0.16 for the temperature and emission measure, respectively).  Note that
there is one sight line that features in both our dataset (sight line 12) and in \citet[their
  sight line 1]{yoshino09}, on which the measured halo temperatures are $(1.68^{+0.30}_{-0.25})
\times 10^6$ and $(2.58^{+1.29}_{-0.79}) \times 10^6~\K$, respectively. Although this discrepancy is
rather large, the 90\%\ confidence intervals do overlap. Furthermore, the \citeauthor{yoshino09}\
confidence interval does not seem to include an estimate of the systematic error due to their
assumed foreground model. Thus, for this sight line, the temperature discrepancy is not significant
given the measurement errors.

In addition to comparing the temperatures and emission measures from the \xmm\ and
\suzaku\ datasets, we can use the temperatures and emission measures from \citet{yoshino09} and
\citet{gupta09b} to calculate the intrinsic 0.5--2.0~\kev\ halo surface brightnesses implied by
their best-fit models, for comparison with our surface brightness measurements. For the
\citet{yoshino09} sight lines, we also take into account the non-solar Fe/O and Ne/O ratios (from
their Table~6).  While the halo temperatures and emission measures obtained with \suzaku\ are in
good overall agreement with our measurements, the median surface brightness inferred from the
best-fit \suzaku\ models is significantly higher than the value from our analysis, although only at the
5\%\ level ($2.68 \times 10^{-12}$ versus $1.54 \times 10^{-12}~\flux\ \pdegsq$; Mann-Whitney $U =
735$, $\pvalue = 0.023$ (two-sided)). However, it should be noted that the supersolar Fe/O and Ne/O
ratios on some of \citeauthor{yoshino09}'s sight lines lead to enhanced halo emission at energies
$\ga$0.8~\kev\ (see their Figure~5, and compare with our Figure~\ref{fig:Spectra1}). If there is
such harder emission from the halo, our solar-abundance $1T$ halo models are unable to model it, and
so our models would tend to yield lower total halo surface brightnesses. (We did experiment with
non-solar Fe/O and Ne/O ratios, but found that in general we were unable to obtain reliable results.)

Above, we noted that the presence of the soft proton contamination in the \xmm\ spectra potentially
introduces some uncertainty in the normalization of the halo X-ray emission. In general, our halo
measurements and those obtained with \suzaku\ (which does not suffer from soft proton contamination)
are in reasonable agreement. This suggests that, in practice, soft proton contamination is not a
major source of bias.

It should be noted that our best-fit halo models attribute somewhat less R45 (3/4~\kev) emission to
the Galactic halo than \citepossessive{kuntz00} analysis of the \rosat\ All-Sky Survey. For sight
lines on which emission is detected, our best-fit models typically imply observed halo R45 count
rates of $(\mbox{18--49}) \times 10^{-6}~\rassrate$ ($\mbox{median} = 27 \times
10^{-6}~\rassrate$). In contrast, \citet{kuntz00} inferred an observed halo R45 count rate of $38.6
\times 10^{-6}~\rassrate$ in the vicinity of the northern Galactic pole (their Table~2:
``Remainder'' $-$ ``Stars''). However, the uncertainty on the \citet{kuntz00} halo R45 count rate is
not stated, so we are unable to determine whether or not this discrepancy is significant.

\subsection{Effect of the Foreground Model}
\label{subsec:ForegroundModel}

Here, we consider the choice of the foreground component in the SXRB model as another potential
source of uncertainty in the determination of the halo emission.  The normalization and spectral
shape of the foreground component may affect the emission measure and temperature measured for the
halo plasma. In our analysis, we followed \hskjm, and used shadowing data from the \rosat\ All-Sky
Survey \citep{snowden00} to fix our foreground model. This method requires extrapolating the
foreground model from the \rosat\ 1/4~\kev\ band to the \xmm\ band ($E \ge 0.4~\kev$). If this
extrapolation leads to an inaccurate foreground model in the \xmm\ band, it will bias our
measurements of the halo temperature and emisison measure.

The above-mentioned \suzaku\ measurements used different techniques for estimating the foreground
emission.  \citet{yoshino09} found a tight correlation between the observed \OVII\ and
\OVIII\ intensities in their sample of spectra, with an non-zero ``floor'' to the \OVII\ emission,
leading them to conclude that their spectra included a uniform local component with \OVII\ and
\OVIII\ intensities of $\sim$2 and $\sim$0~\lineunit, respectively. They subsequently used a
foreground model with these oxygen intensities to obtain their halo measurements. \citet{gupta09b},
meanwhile, compared the emission toward and to the side of the shadow cast in the SXRB by MBM~20,
and thus inferred the contributions to the emission originating in front of and behind the shadowing
cloud. The reasonable agreement between our measurements and these \suzaku\ measurements suggests
that, in fact, our choice of foreground model is not adversely affecting our halo measurements.

\citet{henley13b} adopt a novel, Bayesian approach to inferring the halo X-ray emission from
the \hs\ \xmm\ SXRB survey. They first use the observed time-variation of the oxygen intensities in
directions that have been observed multiple times to specify the prior probability distribution for
the time-variable SWCX intensity in an arbitrary \xmm\ observation. They then combine this prior with
oxygen intensities from other directions to constrain the posterior probability distribution for the
intrinsic halo emission. This new technique yields combined \OVII\ + \OVIII\ halo surface
brightnesses that are generally in reasonable agreement with those inferred from the best-fit halo
models in this paper. However, the halo temperatures inferred from the \OVIII/\OVII\ ratios are
typically $\sim$$0.4 \times 10^6~\K$ lower than those in this paper.

When \citet{henley13b} compare the observations that are included both in that analysis and in the
current analysis, they find that the \citet{henley13b} analysis tends to attribute more
\OVIII\ emission to the foreground than our current foreground model (our foreground model, with $T
= 1.2 \times 10^6~\K$ (Section~\ref{subsubsec:ForegroundModel}), produces virtually no
\OVIII\ emission, which is consistent with \citepossessive{yoshino09} foreground model). This
results in less \OVIII\ emission being attributed to the halo, and hence in a lower temperature
being inferred from the \OVIII/\OVII\ ratio.

\citet{henley13b} repeat our spectral analysis with a higher-temperature foreground model, chosen to
better match the foreground \OVIII\ intensities inferred from their analysis. This revised
foreground model results in halo temperatures that are lower than those presented here, and in better
agreement with those inferred from the \citet{henley13b} \OVIII/\OVII\ ratios. This may imply that
we (and \citet{yoshino09} and \citet{gupta09b}) are underestimating the foreground \OVIII\ intensity
in our spectral fitting, and thus overestimating the halo temperature. However, \citet{henley13b}
also point out that the halo emission likely originates from plasma with a range of temperatures and
in a range of ionization states. If this is the case, the spectral fitting described here will not
necessarily arrive at the same best-fit halo temperature as that inferred from the
\OVIII/\OVII\ ratio alone.

In summary, after comparing our results with those from other studies which use a variety of methods
for determining the foreground emission \citep{yoshino09,gupta09b,henley13b}, we conclude that our
choice of foreground model is not seriously biasing our measurements of the halo surface
brightness. Similarly, our halo temperatures agree with those from other studies that use spectral
fitting \citep{yoshino09,gupta09b}.  However, our temperatures are higher than those inferred from
the halo \OVIII/\OVII\ ratios determined using a novel, Bayesian approach to constraining the halo
emission \citep{henley13b}. While this discrepancy may in part be due to our underestimating the
foreground \OVIII\ intensity, it may also be due to the fact that the halo likely has a complicated
temperature and ionization structure, so different methods of characterizing the halo emission may
yield different temperatures.  Despite this, such temperature measurements are still useful for
testing halo models, provided the predicted halo emission is treated like the true halo
emission. This involves creating synthetic observations from the predicted halo spectra, which are
then analyzed with the same SXRB model as the real observations (\hskjm).

\subsection{Possible Contamination of the Halo Emission}
\label{subsec:Contamination}

In Section~\ref{subsec:ComparisonWithPreviousStudies}, we argued that soft proton contamination was
unlikely to be significantly biasing our results. This conclusion is supported by the fact that we
find no correlation between our measured halo parameters and the level of soft proton contamination,
as quantified by the ratio of the total 2--5~\kev\ model flux to that expected from the
extragalactic background, $F_\mathrm{total}^{2-5} / F_\mathrm{exgal}^{2-5}$ (introduced in
\hsten). For this purpose, we used \hs's measurements of $F_\mathrm{total}^{2-5} /
F_\mathrm{exgal}^{2-5}$, from their Table~2.

SWCX emission is also unlikely to be adversely affecting our halo measurements: our observations
were selected from \hs's catalog as they were expected to be the least contaminated by SWCX emission
(Section~\ref{subsec:ObservationSelection}), and in Section~\ref{subsec:ForegroundModel} we argued
that our choice of foreground model is not seriously biasing our measurements of the halo surface
brightness. In this subsection we consider other potential sources of contamination.

Our halo measurements are unlikely to be contaminated by emission from the original targets of the
\xmm\ observations. In \hs, we experimented with increasing the radii of the exclusion regions used
to excise the \xmm\ target objects, and concluded that the SXRB \OVII\ and \OVIII\ intensities were
not being contaminated by emission from those targets (see Section~3.6 of \hs). Since the \OVII\
and \OVIII\ emission dominates the halo emission in the \xmm\ band, the results from \hs\ imply that
our halo measurements are not contaminated by emission from the \xmm\ targets.

\begin{figure}
\plotone{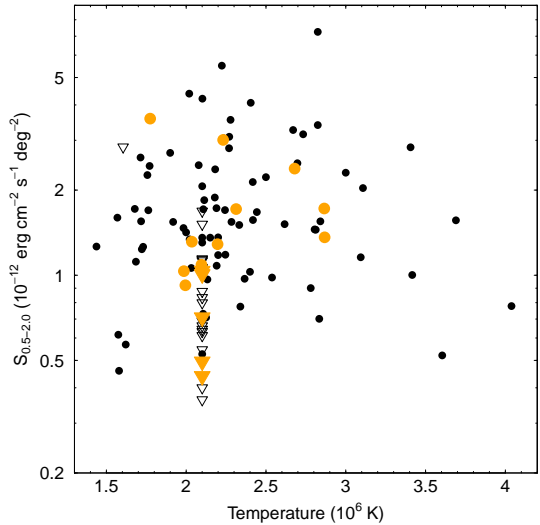}
\caption{0.5--2.0~\kev\ halo surface brightnesses against halo temperatures. Sight lines on which
  the original target of the \xmm\ observation was a galaxy cluster are colored orange. The triangles
  indicate upper limits on the surface brightnesses.
  \label{fig:ClustervNonCluster}}
\end{figure}

In Figure~\ref{fig:ClustervNonCluster}, we compare the results from sight lines on which the target
object was a galaxy cluster (orange) with those from other sight lines (black). We single out galaxy
clusters because it can be difficult to determine the extent of the cluster emission from a visual
inspection of the X-ray images, and so emission from the cluster periphery could potentially
contaminate our halo measurements. However, Figure~\ref{fig:ClustervNonCluster} shows that the halo
surface brightnesses measured on cluster sight lines are not systematically higher than those on
other sight lines.  For sight lines on which halo emission is detected, the median surface
brightnesses on cluster and non-cluster sight lines are $1.36 \times 10^{-12}$ and $1.55 \times
10^{-12}~\flux\ \pdegsq$, respectively. We therefore conclude that emission from the peripheries of
targeted galaxy clusters is not contaminating our halo measurements.

Our halo measurements could also potentially be contaminated by non-targeted galaxy groups or
clusters that happen to lie in the \xmm\ fields of view. Unless such objects are particularly
bright, they may have escaped being noticed in our visual inspection of the X-ray images. Indeed, in
Section~\ref{subsec:DataReduction}, we noted that when we used the exclusion regions used in
\hsten\ and \hs, our initial spectral fitting yielded unusually high halo temperatures on nine sight
lines ($T \sim \tenMK$). Upon re-examination of these sight lines, using newly created adaptively
smoothed, QPB-subtracted images, we found additional diffuse emission that had not been adequately
removed after the initial visual inspection. After removing these additional regions of diffuse
emission, the spectral fitting generally yielded typical halo temperatures for these sight lines ($T
\sim \twoMK$).

\begin{figure}
\plotone{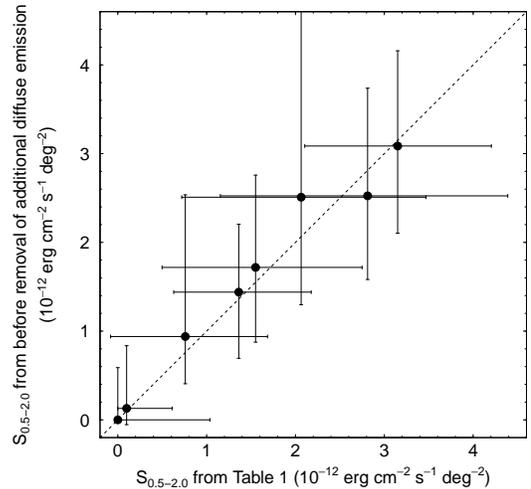}
\caption{0.5--2.0~\kev\ surface brightnesses of the $\sim$\twoMK\ halo emission measured before
  (ordinates) and after (abscissae) removing regions of diffuse emission that had not been removed
  in \hs\ (see Section~\ref{subsec:DataReduction}). We have plotted the results for the sight lines
  indicated by a ``d'' in Column~1 of Table~\ref{tab:FitResults} (apart from sight line 83). These
  sight lines yielded a halo temperature of $\sim$\tenMK\ when we used the \hs\ exclusion regions,
  but a temperature of $\sim$\twoMK\ after removing additional regions of diffuse emission. The
  ordinates were obtained by analyzing the originally extracted SXRB spectra with a $2T$ model to
  model the non-foreground, non-extragalactic-background emission -- the plotted values are for the
  cooler component of this model.
  \label{fig:OldvNewSB}}
\end{figure}

It is likely that similar regions of diffuse emission lie in the other fields of view, which we have
not re-examined. However, such regions are unlikely to be contaminating our halo measurements.
Before removing the additional regions of diffuse emission from the nine sight lines noted above, we
analyzed the SXRB spectra using a $2T$ model to model the non-foreground,
non-extragalactic-background emission: one component to model the $\sim$\twoMK\ halo emission, and a
$\sim$\tenMK\ component of uncertain origin. The properties of the cooler component from before the
removal of the additional regions of diffuse emission and the properties of the halo component from
after the removal of said regions are consistent with each other. In particular, the surface
brightnesses of the $\sim$\twoMK\ emission were not greatly affected (see
Figure~\ref{fig:OldvNewSB}). We therefore conclude that non-targeted extragalactic diffuse emission
that happens to lie in our remaining \xmm\ fields of view is not contaminating our halo
measurements. At least in part, this is likely because galaxy groups and clusters tend to be much
hotter than $\sim$\twoMK\ \citep[e.g.,][]{osmond04,sanderson03}.

Finally, we consider contamination from point sources. The combined flux from point sources below
our source removal flux threshold ($\Fsource = 1 \times 10^{-14}~\flux$) is modeled by our
extragalactic background model (Section~\ref{subsubsec:ExtragalacticModel}). Our estimates of the
systematic errors take into account the uncertainty in the normalization of the extragalactic
background (Section~\ref{subsec:Systematic}). Although the uncertainty in the normalization of the
extragalactic background is $\pm 1.1 \times 10^{-12}~\flux\ \pdegsq$
(Section~\ref{subsec:Systematic}), we find that this leads to an uncertainty in the halo surface
brightness typically of only $\pm 0.3 \times 10^{-12}~\flux\ \pdegsq$. Somewhat surprisingly,
increasing the brightness of the extragalactic background from its nominal value generally leads to
an increase in the halo surface brightness. This is because increasing the brightness of the
extragalactic background causes the soft proton component to decrease in brightness to compensate,
and the combination of these effects leads to more flux being attributed to the halo emission (see
Section~\ref{subsec:ComparisonWithPreviousStudies}).

Sources brighter than the above threshold were excised in the automated source removal
(Section~\ref{subsec:DataReduction}), using circles of radius 50\arcsec, which enclose
$\approx$90\%\ of each source's flux.\footnote{This encircled energy fraction, which depends only
  weakly on the off-axis angle, was calculated from the best-fit King profiles to the MOS
  telescopes' point spread functions, using the \texttt{XRT1\_XPSF\_0014.CCF} and
  \texttt{XRT2\_XPSF\_0014.CCF} calibration files (see the \xmm\ Calibration Access and Data
  Handbook;
  http://xmm.vilspa.esa.es/external/xmm\_sw\_cal/calib/\\documentation/CALHB/node30.html).}  (The
brighter sources were removed with larger source exclusion regions, positioned by eye over the
sources.) Using the source fluxes either from the \xmm\ Serendipitous Source Catalog or from the
source detection that we ran ourselves, we can estimate the total flux in each field due to photons
from automatically removed sources that have spilled out of the source exclusion regions.
Typically, this flux is $<$$0.5 \times 10^{-12}~\flux\ \pdegsq$. This is less than the uncertainty
in the normalization of the extragalactic background, and the effect of this contaminating flux on
the halo surface brightnesses is likely to be small (as noted above, the uncertainty in the
extragalactic normalization leads to an uncertainty in the halo surface brightness typically of only
$\pm 0.3 \times 10^{-12}~\flux\ \pdegsq$).

In summary, our halo measurements are unlikely to be significantly affected by contamination from the
original \xmm\ targets, from non-targeted group or clusters of galaxies lying in the fields of view,
or from photons spilling out of the source exclusion regions defined in the automated source
removal.

\subsection{The Morphology of the Hot Galactic Halo}
\label{subsec:Morphology}

While the halo temperature appears to be fairly uniform, the halo emission measure exhibits
considerable variation across the sky (Figure~\ref{fig:Maps}). In Section~\ref{subsec:LBVariation},
we showed that the halo emission measure tends to increase from the outer to the inner Galaxy, at
least in the southern Galactic hemisphere. We previously noted a similar trend for the
\textit{observed} \OVII\ and \OVIII\ intensities in the south (\hs). Since the SWCX intensity is not
expected to be correlated with Galactic longitude, we argued in \hs\ (Section~4.3.1) that the
observed trend reflected an increase in the halo's \textit{intrinsic} emission from the outer to the
inner Galaxy, in agreement with the halo measurements presented here. These results argue against
the hot halo having a simple plane-parallel disk-like morphology, in which case the emission measure
would be independent of Galactic longitude. Instead, these results suggest a halo that is
concentrated toward the Galactic Center.

The variation (or lack thereof) of the halo emission measure with latitude is also different
from that expected for a plane-parallel disk-like halo morphology. In such a morphology,
the emission measure is expected to decrease with latitude as $1/\sin |b|$. Instead, we
find the halo emission measure to weakly increase with latitude in the north, and to be uncorrelated
with latitude in the south (see Table~\ref{tab:CorrelationCoefficients}, and
Figures~\ref{fig:Correlation}(a) and (b)). Similarly, we previously found that our SXRB oxygen
intensity measurements argued against a plane-parallel halo model in the northern Galactic
hemisphere (see \hs, Section~4.3.3).

\citet{yoshino09} did find that the halo emission measure decreased with latitude. However, the
decrease with latitude is steeper than that expected for a plane-parallel model (see their Figure~7;
$\mbox{E.M.} \times \sin |b|$ is expected to be constant for a plane-parallel model).  Note also
that the \citet{yoshino09} dataset contains far fewer sight lines than ours. The fact that our halo
emission measures do not decrease with increasing latitude, contrary to what is expected for a
disk-like halo morphology, and contrary to the \citet{yoshino09} \suzaku\ results, is unlikely to be
due to soft proton contamination (a problem from which \suzaku\ does suffer). We argued in
Sections~\ref{subsec:ComparisonWithPreviousStudies} and \ref{subsec:Contamination}, above, that such
contamination does not seriously bias our halo emission measures.

The fact that our halo emission measures do not decrease with increasing latitude is also
unlikely to be due to our using \HI\ column densities to attenuate the halo emission. By using such
column densities, we could potentially be neglecting the contributions to the absorption from
regions containing \HII\ or molecular H.  For each sight line, we extracted the interstellar
reddening, $E(B-V)$, from the \citet{schlegel98} maps (derived from \textit{COBE}/DIRBE-corrected
\iras\ data), and converted to \NH\ using the conversion relation from \citet{guver09}. This
relation was derived from X-ray spectral analysis of supernova remnants, and hence should yield the
total hydrogen column density (although note that the hydrogen column densities in that
study are typically larger than those used in the present paper).  Surprisingly, the hydrogen
column densities obtained in this way were typically \textit{smaller} than the \HI\ column
densities. Furthermore, the effect on the attenuation factor ($\exp(-\sigma\NH)$, where $\sigma$ is
the photoelectric absorption cross-section) was typically less than a few percent in the vicinity of
the oxygen lines, and uncorrelated with latitude. Hence our using \HI\ column densities is unlikely
to be responsible for our emission measure measurements not following the trend expected
for a disk-like halo morphology. Note also that \citet{yoshino09} used \HI\ column densities for
all but two of their sight lines.

In summary, our observations suggest that the halo is concentrated toward the Galactic Center.
Other than that, the morphology of the hot halo remains uncertain. However, our observed
emission measures do not vary with latitude in the way expected for a disk-like halo
morphology. The patchiness of the halo emission makes it difficult to determine the halo's
underlying global morphology.

\subsection{The Origin of the Hot Galactic Halo}
\label{subsec:Origins}

We defer a detailed comparison of our observations to models of the hot halo to a follow-up paper
(D.~B. Henley et~al., 2013, in preparation). However, here we make some general comments on the
implications of our results for the origin of the hot halo.

We cannot use energy arguments to distinguish between SN-driven outflows and accretion of
extragalactic material as the sources of the hot halo plasma, since both processes have more than
enough energy to power the X-ray emission. The latitude-corrected intrinsic halo surface brightness,
$\Stotal \sin |b|$, implies a 0.5--2.0~\kev\ halo luminosity of $8 \pi \Stotal \sin |b| = 8.4 \times
10^{35}~\ergps\ \kpc^{-2}$ (where we have used the median of $\Stotal \sin |b|$, $1.07 \times
10^{-12}~\flux\ \pdegsq$, including non-detections at their best-fit values). As has previously been
noted, this is much less than the energy available from SNe ($8 \times 10^{38}~\ergps\ \kpc^{-2}$;
e.g., \citealt{yao09}). It is also much less than the energy potentially available from
accretion. The Galactic escape speed in the vicinity of the Sun (540~\kmps; \citealt{smith07c})
implies a gravitational potential of $-1.5 \times 10^{15}~\erg\ \gram^{-1}$.  If the Galaxy accretes
mass at a rate of $\sim$0.4~\Msolpy\ \citep{chiappini09} over a disk of radius $\sim$20~\kpc,
$\sim$$3 \times 10^{37}~\ergps\ \kpc^{-2}$ is available from accretion in the vicinity of the
Sun.\footnote{Of course, the hot gas observed in the halo has not fallen all the way to the
  Sun. However, integrating the vertical gravitational acceleration at the solar circle
  \citep{ferriere98a} to a height of 3~\kpc, we find that the estimate of the available energy need
  only be revised downward by a few percent for hot gas a few \kpc\ above the disk.} Detailed models
of the outflow and accretion scenarios are needed to determine how much of the available energy is
actually radiated as X-rays.

In Section~\ref{subsec:Morphology}, we pointed out that our and \hs's measurements suggest a halo
that is concentrated toward the Galactic Center. Such a morphology may be due to a halo of accreted
material centered on the Galactic Center, or to an increase in the Galactic SN rate toward the inner
Galaxy (e.g., the SN rate per unit area increases by a factor of $\sim$20 from 2~\kpc\ outside the
solar circle to 2~\kpc\ within the solar circle; \citealt{ferriere98a}).

We noted in Section~\ref{sec:Results} that the halo emission measure and intrinsic surface
brightness vary by an order of magnitude over the sky, while the temperature is fairly uniform
(see Figure~\ref{fig:EMvT}). Figures~\ref{fig:Maps}(b) and \ref{fig:Maps}(d) show that the halo
emission is patchy (\citealt{yoshino09}; \hsten; \hskjm). Such patchiness may favor a stochastic,
inhomogeneous energy source, such as SNe, as the source of the hot halo (\hskjm). However, if
accreting extragalactic material fragments as it accretes, it too could lead to patchy emission.

In summary, arguments can be made in favor of both the outflow and the accretion scenarios based on
our current set of observations. Detailed models are needed to determine which of these processes
dominates the X-ray emission, or if both processes play a significant role. Not only should such
models match the observed X-ray temperature and surface brightness, ideally they should also explain
the observed variation in the halo brightness -- both the general increase from the inner to the outer
Galaxy, and the patchiness of the emission.

\section{SUMMARY}
\label{sec:Summary}

We have presented measurements of the Galactic halo's X-ray emission on 110 \xmm\ sight lines. This
is an approximately fourfold increase in the number of sight lines over the previous largest study of
the Galactic halo with CCD-resolution X-ray spectra (\hskjm). Our sample of observations is drawn
from an \xmm\ survey of the SXRB (\hs). We selected these observations as they should be the least
contaminated by charge exchange emission from within the solar system. We analyzed the spectra with
a standard SXRB model, with components representing the foreground, Galactic halo, and extragalactic
background emission.

We detected emission from $\sim$\twoMK\ plasma on 87 of our sight lines (79\%), with typical
emission measures of $(\mbox{1.4--3.0}) \times 10^{-3}~\emismeas$, and typical intrinsic
0.5--2.0~\kev\ surface brightnesses of $(\mbox{1.1--2.3}) \times 10^{-12}~\flux\ \pdegsq$
(Section~\ref{sec:Results}). The halo emission measure tends to increase from the outer to the inner
Galaxy in the southern Galactic hemisphere (Section~\ref{subsec:LBVariation}). There is some
evidence that the halo is hotter and has a larger emission measure in the southern hemisphere than
in the north (Section~\ref{subsec:HemisphereDifferences}). However, the differences may be partly
due to the fact that we are not comparing equivalent regions in both hemispheres. Because of the
presence of the Sco-Cen superbubble, we excluded the region toward the inner Galaxy in the northern
hemisphere but not in the south, and, as noted above, the emission measure increases toward the
inner Galaxy in the south. In addition, it should be noted that the difference in the median
temperature between the hemispheres ($\sim$$0.2 \times 10^6~\K$) is less than the typical error on
the temperature ($\sim \pm 0.4 \times 10^6~\K$).

Our halo emission measures and surface brightnesses are in reasonable agreement with those measured
with \suzaku\ \citep{yoshino09,gupta09b}
(Section~\ref{subsec:ComparisonWithPreviousStudies}). Similarly, the halo surface brightnesses
attributable to the oxygen \Kalpha\ lines (derived from our best-fit halo models) are generally in
reasonable agreement with those derived from a Bayesian analysis of the oxygen intensities from the
HS12 \xmm\ SXRB survey \citep{henley13b} (Section~\ref{subsec:ForegroundModel}). Since these studies
all used different methods for estimating the foreground emission, and since \suzaku\ does not
suffer from significant soft proton contamination, we conclude that neither our choice of foreground
model nor soft proton contamination significantly bias our halo surface brightness measurements.
Contamination from the original targets of the \xmm\ observations, from non-targeted galaxy groups
and clusters lying in the fields of view, and from point sources are also not significantlly
affecting our measurements (Section~\ref{subsec:Contamination}).

Our halo temperatures are in agreement with those measured with \suzaku, but are higher than those
inferred from the halo \OVIII/\OVII\ ratios determined from the above-mentioned Bayesian analysis
of the SXRB oxygen lines \citep{henley13b}. In Section~\ref{subsec:ForegroundModel}, we mentioned
two possible explanations for this discrepancy.  The discrepancy may indicate that we (and the
\suzaku\ studies) are underestimating the foreground \OVIII\ intensity, thus biasing the halo
measurements.  Alternatively, the discrepancy may be due to the halo having a complex
multitemperature, multi-ionization-state structure, meaning that the \OVIII/\OVII\ ratio and the
broadband spectral analysis will not necessarily yield the same best-fit temperature.

Our halo emission measures do not decrease with increasing Galactic latitude, contrary to
what is expected for a plane-parallel disk-like halo morphology. This result appears not
to be an artifact of soft proton contamination, nor of our using \HI\ column densities when
calculating the attenuation of the halo emission. The details of the morphology of the halo remain
uncertain, but the variation of the emission measure with longitude suggests that the halo is
concentrated toward the Galactic Center (Section~\ref{subsec:Morphology}).

We are unable to distinguish between extragalactic accretion and outflows from the disk as the
source of the $\sim$\twoMK\ halo plasma. Both processes have more than enough energy to maintain the
halo's surface brightness, and other aspects of the halo emission (such as the increase in emission
measure toward the inner Galaxy and the general patchiness of the emission) could plausibly be
explained by either scenario (Section~\ref{subsec:Origins}). A detailed comparison of our
measurements with the predictions of hydrodynamical models of the halo is needed to distinguish
between different scenarios for the heating of the halo.  We will carry out such a comparison in a
follow-up paper (D.~B. Henley et~al. 2013, in preparation).

\acknowledgements

We thank the anonymous referee, whose comments helped significantly improve this paper.
This research is based on observations obtained with \xmm, an ESA science mission with instruments
and contributions directly funded by ESA Member States and NASA.
We acknowledge use of the R software package \citep{R}.
This research has made use of the SIMBAD database, operated at CDS, Strasbourg, France, and of the
NASA/IPAC Extragalactic Database (NED) which is operated by the Jet Propulsion Laboratory,
California Institute of Technology, under contract with the National Aeronautics and Space
Administration.
This research was funded by NASA grant NNX08AJ47G, awarded through the Astrophysics Data Analysis
Program.

\bibliography{references}

\clearpage
\setcounter{table}{\value{ResultsTable}}
{
  \LongTables
  \tabletypesize{\scriptsize}
  \setlength{\tabcolsep}{2.5pt}
  \begin{landscape}
    \input{tab1}
    \clearpage
  \end{landscape}
}

\end{document}

%% file: newcom.tex
\newcommand{\HI}{\ion{H}{1}}
\newcommand{\HII}{\ion{H}{2}}

\newcommand{\OVII}{\ion{O}{7}}
\newcommand{\OVIII}{\ion{O}{8}}

\newcommand{\Kalpha}{K$\alpha$}

\newcommand{\Msol}{\ensuremath{M_{\odot}}}
\newcommand{\Ne}{\ensuremath{n_{\mathrm{e}}}}

\newcommand{\NH}{\ensuremath{N_{\mathrm{H}}}}

\newcommand{\nm}{\ensuremath{\mbox{\nm}}}
\newcommand{\cm}{\ensuremath{\mbox{cm}}}

\newcommand{\km}{\ensuremath{\mbox{km}}}

\newcommand{\pc}{\ensuremath{\mbox{pc}}}
\newcommand{\kpc}{\ensuremath{\mbox{kpc}}}

\newcommand{\s}{\ensuremath{\mbox{s}}}

\newcommand{\yr}{\ensuremath{\mbox{yr}}}

\newcommand{\gram}{\ensuremath{\mbox{g}}}

\newcommand{\kev}{\ensuremath{\mbox{keV}}}

\newcommand{\erg}{\ensuremath{\mbox{erg}}}

\newcommand{\sr}{\ensuremath{\mbox{sr}}}

\newcommand{\K}{\ensuremath{\mbox{K}}}

\newcommand{\ph}{\ensuremath{\mbox{photons}}}
\newcommand{\counts}{\ensuremath{\mbox{counts}}}

\newcommand{\parcminsq}{\ensuremath{\mbox{arcmin}^{-2}}}
\newcommand{\pdegsq}{\ensuremath{\mbox{deg}^{-2}}}

\newcommand{\pcmsq}{\ensuremath{\cm^{-2}}}

\newcommand{\ps}{\ensuremath{\s^{-1}}}
\newcommand{\psr}{\ensuremath{\sr^{-1}}}
\newcommand{\pyr}{\ensuremath{\yr^{-1}}}

\newcommand{\emismeas}{\ensuremath{\cm^{-6}}\ \pc}
\newcommand{\ergps}{\erg\ \ps}
\newcommand{\flux}{\erg\ \pcmsq\ \ps}
\newcommand{\lineunit}{\ph\ \pcmsq\ \ps\ \psr}

\newcommand{\kmps}{\km\ \ps}
\newcommand{\Msolpy}{\Msol\ \pyr}
\newcommand{\pownorm}{\ph\ \pcmsq\ \ps\ \psr\ \ensuremath{\kev^{-1}}}

\newcommand{\rassrate}{\counts\ \ps\ \parcminsq}

\newcommand{\chandra}{\textit{Chandra}}

\newcommand{\iras}{\textit{IRAS}}

\newcommand{\rosat}{\textit{ROSAT}}

\newcommand{\suzaku}{\textit{Suzaku}}

\newcommand{\xmm}{\textit{XMM-Newton}}

\newcommand{\citepossessive}[1]{\citeauthor{#1}'s \citeyearpar{#1}}

\newcommand{\chisq}{\ensuremath{\chi^2}}

\newcommand{\raymondsmith}{\citeauthor{raymond77} (\citeyear{raymond77} and updates)}
\providecommand{\eqref}[1]{equation~(\ref{#1})}

\newcommand{\esas}{\textit{XMM}-ESAS}

%% file: tab2.tex
{
\setlength{\tabcolsep}{3.5pt}
\begin{deluxetable*}{clcccccccccccccc}
\tablewidth{0pt}
\tablecaption{Medians and Quartiles of the Halo Temperature, Emission Measure, and Surface Brightness\label{tab:Quartiles}}
\tablehead{
		&			& \multicolumn{4}{c}{Temperature}				&& \multicolumn{4}{c}{Emission Measure}					&& \multicolumn{4}{c}{\Stotal\tablenotemark{a}} \\
		&			& \multicolumn{4}{c}{($10^6~\K$)}				&& \multicolumn{4}{c}{($10^{-3}~\emismeas$)}				&& \\
\cline{3-6} \cline{8-11} \cline{13-16} \\
\colhead{Row}	& \colhead{Data subset}	& \colhead{$N$}	& \colhead{LQ}	& \colhead{Med}	& \colhead{UQ}	&& \colhead{$N$}	& \colhead{LQ}	& \colhead{Med}	& \colhead{UQ}	&& \colhead{$N$}	& \colhead{LQ}	& \colhead{Med}	& \colhead{UQ}
}
\startdata
 1 & Full dataset -- detections\tablenotemark{b}                                 &       83 &     2.01 &     2.22 &     2.64 &&       87 &     1.38 &     1.91 &     3.04 &&       87 &     1.14 &     1.54 &    2.34  \\
 2 & Full dataset -- non-detections (upper limits)                               &  \nodata &  \nodata &  \nodata &  \nodata &&       23 &     0.81 &     1.04 &     1.43 &&       23 &     0.62 &     0.80 &    1.09  \\
 3 & Northern Galactic hemisphere\tablenotemark{c}                               &       45 &     1.76 &     2.13 &     2.37 &&       66 &     0.52 &     1.41 &     2.12 &&       66 &     0.44 &     1.07 &    1.56  \\
 4 & Southern Galactic hemisphere\tablenotemark{c}                               &       38 &     2.13 &     2.30 &     2.72 &&       44 &     1.27 &     2.05 &     3.32 &&       44 &     1.18 &     1.71 &    3.06  \\
 5 & Southern Galactic hemisphere ($|l| > 50\degr$)\tablenotemark{c,d}           &       27 &     2.11 &     2.28 &     2.71 &&       33 &     1.14 &     1.49 &     2.53 &&       33 &     1.03 &     1.45 &    2.43 
\enddata

\tablecomments{For each quantity, we tabulate the number of sight lines ($N$), the lower quartile
  (LQ), the median (Med), and the upper quartile (UQ). The results are taken from Columns~12, 13,
  and 15 of Table~\ref{tab:FitResults}.}

\tablenotetext{a}{Intrinsic 0.5--2.0~\kev\ surface brightness in $10^{-12}~\flux~\pdegsq$.}
\tablenotetext{b}{When calculating the quartiles of the temperatures, we exclude sight lines for
  which the temperature is fixed at $2.1 \times 10^6~\K$.}
\tablenotetext{c}{When calculating the quartiles of the temperatures, we exclude non-detections and
  sight lines for which the temperature is fixed at $2.1 \times 10^6~\K$. When calculating the
  quartiles of the emission measures, we use the best-fit emission measures from
  Table~\ref{tab:FitResults} for all sight lines, including those that
  yield non-detections (and similarly for the surface brightnesses).}
\tablenotetext{d}{See Equation~(\ref{eq:lprime}) for the definition of $|l|$.}

\end{deluxetable*}
}

%% file: tab3.tex
\begin{deluxetable*}{llcccc}
\tablewidth{0pt}
\tablecaption{Correlation Coefficients for Halo Temperature and Emission Measure against Galactic Longitude and Latitude\label{tab:CorrelationCoefficients}}
\tablehead{
	&				& \multicolumn{2}{c}{Temperature}						& \multicolumn{2}{c}{Emission Measure} \\
	&				& \colhead{$\tau$\tablenotemark{a}}	& \colhead{$p$ value\tablenotemark{b}}	& \colhead{$\tau$\tablenotemark{a}}	& \colhead{$p$ value\tablenotemark{b}}
}
\startdata
North   & $|l|$\tablenotemark{c}        & $-0.07$ ($-0.27$, $+0.15$)            & 0.51                                  & $-0.17$ ($-0.28$, $-0.03$)            & 0.050  \\
        & $|b|$                         & $-0.13$ ($-0.30$, $+0.05$)            & 0.21                                  & $+0.18$ ($+0.02$, $+0.29$)            & 0.033  \\
South   & $|l|$\tablenotemark{c}        & $-0.22$ ($-0.42$, $-0.02$)            & 0.055                                 & $-0.44$ ($-0.55$, $-0.34$)            & $4.0 \times 10^{-5}$  \\
        & $|b|$                         & $-0.05$ ($-0.27$, $+0.19$)            & 0.65                                  & $+0.02$ ($-0.18$, $+0.22$)            & 0.82 
\enddata
\tablenotetext{a}{Kendall's $\tau$ \citep[e.g.,][]{press92}, with the 90\% bootstrap confidence interval shown in parentheses.}
\tablenotetext{b}{Probability of observing a correlation coefficient at least as extreme as the value that is observed, under the null hypothesis of there being no correlation.}
\tablenotetext{c}{See Equation~(\ref{eq:lprime}).}
\end{deluxetable*}

%% file: tab1.tex
\begin{deluxetable}{l@{\hskip -1cm}llccccccccccr@{/}lc}
\tablewidth{0pt}
\tabletypesize{\scriptsize}
\tablecaption{Observation Details and Spectral Fit Results\label{tab:FitResults}}
\tablehead{
\colhead{Sight}		& \colhead{Obs. Id}	& \colhead{Target}	& \colhead{$l$}		& \colhead{$b$}		& \colhead{$t^\mathrm{exp}_1$}	& \colhead{$\Omega_1$}		& \colhead{$t^\mathrm{exp}_2$}	& \colhead{$\Omega_2$}		& \colhead{F/g R12}			& \colhead{\NH}			& \colhead{Halo $T$}		& \colhead{Halo E.M.}			& \multicolumn{2}{c}{$\chisq/\mathrm{dof}$}	& \colhead{\Stotal\tablenotemark{b}} \\
\colhead{line}		&			&			& \colhead{(deg)}	& \colhead{(deg)}	& \colhead{(ks)}		& \colhead{(arcmin$^2$)}	& \colhead{(ks)}		& \colhead{(arcmin$^2$)}	& \colhead{rate\tablenotemark{a}}	& \colhead{($10^{20}~\pcmsq$)}	& \colhead{($10^6~\K$)}		& \colhead{($10^{-3}~\emismeas$)}	& \multicolumn{2}{c}{}				& \colhead{} \\
\colhead{(1)}		& \colhead{(2)}		& \colhead{(3)}		& \colhead{(4)}		& \colhead{(5)}		& \colhead{(6)}			& \colhead{(7)}			& \colhead{(8)}			& \colhead{(9)}			& \colhead{(10)}			& \colhead{(11)}		& \colhead{(12)}		& \colhead{(13)}			& \multicolumn{2}{c}{(14)}			& \colhead{(15)}
}
\startdata
1                      &   0302580501 &         RX J2359.5$-$3211 &   5.676 & $-77.683$ &            20.0 &             378 &            20.7 &             462 &        440 &       1.18 &               $1.77^{+0.15}_{-0.11} \pm 0.05$ &      $6.76^{+1.23}_{-1.18}\,^{+0.90}_{-0.86}$ &        513.03 &           434 &   $3.58^{+0.81}_{-0.77}$ \\
2                      &   0550460801 &                   IC 5179 &   6.479 & $-55.914$ &            24.9 &             373 &            25.4 &             454 &        652 &       1.39 &      $2.42^{+0.30}_{-0.19}\,^{+0.32}_{-0.11}$ &      $2.23^{+0.64}_{-0.63}\,^{+0.56}_{-0.58}$ &        484.98 &           541 &   $2.14^{+0.81}_{-0.82}$ \\
3                      &   0304080501 &              H 2356$-$309 &  12.896 & $-78.051$ &            16.9 &             353 &            17.0 &             370 &        426 &       1.44 &               $2.10^{+0.11}_{-0.10} \pm 0.05$ &      $5.49^{+0.85}_{-0.79}\,^{+0.61}_{-0.74}$ &        391.39 &           355 &   $4.21^{+0.80}_{-0.83}$ \\
4                      &   0556210301 &  SDSS J163408.64+331242.0 &  54.428 & $+42.051$ &            10.4 &             465 &            11.3 &             475 &        441 &       1.70 &      $3.42^{+6.36}_{-1.07}\,^{+4.64}_{-0.65}$ &      $0.77^{+0.84}_{-0.56}\,^{+0.42}_{-0.37}$ &        357.93 &           384 &   $1.00^{+1.21}_{-0.86}$ \\
5                      &   0505880101 &  SDSS J163246.54+340526.1 &  55.564 & $+42.457$ &            13.5 &             342 &            14.6 &             492 &        447 &       1.73 &      $2.11^{+0.23}_{-0.28}\,^{+0.09}_{-0.07}$ &               $2.38^{+0.88}_{-0.83} \pm 0.58$ &        324.86 &           380 &   $1.84^{+0.81}_{-0.79}$ \\
\enddata
\end{deluxetable}